\documentclass[twocolumn]{aastex7}

\usepackage{xcolor}
\usepackage{xspace}
\usepackage{amsmath}
\usepackage{natbib}
\usepackage{bm}
\usepackage{tablefootnote}

\defcitealias{Son2025}{S25}
\defcitealias{Chung2025}{C25}
\defcitealias{Wiseman2026}{W26}
\defcitealias{Behroozi2013}{B13}
\defcitealias{Madau2014}{MD14}
\newcommand{\Son}{\citetalias{Son2025}\xspace}
\newcommand{\Chung}{\citetalias{Chung2025}\xspace}
\newcommand{\Wiseman}{\citetalias{Wiseman2026}\xspace}
\newcommand{\Behroozi}{\citetalias{Behroozi2013}\xspace}
\newcommand{\Madau}{\citetalias{Madau2014}\xspace}

\shorttitle{Old Universe, Young SNe Ia}
\shortauthors{Murakami \& Tweddle et al.}

\graphicspath{{./}{figures/}}

\begin{document}

\title{Old Universe, Young SNe Ia: A Statistical Analysis of Type Ia Supernova Progenitor Age from 6,983 TITAN Host Galaxies, and Implications for Cosmology}

\correspondingauthor{Yukei S. Murakami, Jack W. Tweddle}

\author[0000-0002-8342-3804]{Yukei S. Murakami}
\email[show]{ymuraka2@jhu.edu}
\affiliation{Department of Physics and Astronomy, Johns Hopkins University, Baltimore, MD 21218, USA}

\author[0009-0004-5681-545X]{Jack W. Tweddle}
\email[show]{jack.tweddle@physics.ox.ac.uk}
\affiliation{Astrophysics sub-Department, Department of Physics, University of Oxford, Keble Road, Oxford, OX1 3RH, UK}

\author[0000-0002-3073-1512]{Phil Wiseman}
\email{}
\affiliation{School of Physics and Astronomy, University of Southampton, Southampton SO17 1BJ, UK}

\author[0000-0001-8738-6011]{Saurabh~W.~Jha}
\email{}
\affiliation{Department of Physics and Astronomy, Rutgers, the State University of New Jersey, Piscataway, NJ 08854, USA}

\author[0000-0002-6124-1196]{Adam G. Riess}
\email{}
\affiliation{Department of Physics and Astronomy, Johns Hopkins University, Baltimore, MD 21218, USA}
\affiliation{Space Telescope Science Institute, 3700 San Martin Drive, Baltimore, MD 21218, USA}

\author[0000-0002-8229-1731]{Stephen J. Smartt}
\email{}
\affiliation{Astrophysics sub-Department, Department of Physics, University of Oxford, Keble Road, Oxford, OX1 3RH}

\author[0000-0001-8788-1688]{Maria Vincenzi}
\email{}
\affiliation{Astrophysics sub-Department, Department of Physics, University of Oxford, Keble Road, Oxford, OX1 3RH}

\author[0000-0002-5864-1332]{Gautham Adamane Pallathadka}
\email{}
\affiliation{Department of Physics and Astronomy, Johns Hopkins University, Baltimore, MD 21218, USA}

\author[0000-0001-5201-8374]{Dillon Brout}
\email{}
\affiliation{Departments of Astronomy and Physics, Boston University, Boston MA 02215}

\author[0000-0002-6230-0151]{David O. Jones}
\email{}
\affiliation{Institute for Astronomy, University of Hawai'i, 640 N.~A'ohoku Pl., Hilo, HI 96720, USA}

\author[0000-0002-4934-5849]{Daniel Scolnic}
\email{}
\affiliation{Department of Physics, Duke University, Durham, NC 27708, USA}

\author[0009-0003-4631-3184]{Elijah G. Marlin}
\email{}
\affiliation{Departments of Astronomy and Physics, Boston University, Boston MA 02215}

\author[0000-0002-8012-6978]{Brodie A. Popovic}
\email{}
\affiliation{School of Physics and Astronomy, University of Southampton, Southampton SO17 1BJ, UK}

\author[0000-0002-1296-6887]{Lluís Galbany}
\email{}
\affiliation{Institute of Space Sciences (ICE-CSIC), Campus UAB, Carrer de Can Magrans, s/n, E-08193 Barcelona, Spain}
\affiliation{Institut d'Estudis Espacials de Catalunya (IEEC), 08860 Castelldefels (Barcelona), Spain}

\author[0000-0002-8538-9195]{Brian P. Schmidt}
\email{}
\affiliation{Research School of Astronomy and Astrophysics, Australian National University, Canberra, ACT 0200, Australia}

\author[0000-0002-9955-8797]{Keto D. Zhang}
\email{ketoz@ipac.caltech.edu}
\affiliation{IPAC, California Institute of Technology, MC 100-22, 1200 E California Blvd Pasadena, CA 91125, USA}

\author[0000-0003-0928-0494]{Mitchell Dixon}
\email{}
\affiliation{Institute for Astronomy, University of Hawai'i, 640 N.~A'ohoku Pl., Hilo, HI 96720, USA}

\author[0000-0003-2037-4619]{Conor Larison}
\email{}
\affiliation{Space Telescope Science Institute, 3700 San Martin Drive, Baltimore, MD 21218, USA}

\author[0000-0001-7113-2738]{Henry C. Ferguson}
\email{}
\affiliation{Space Telescope Science Institute, 3700 San Martin Drive, Baltimore, MD 21218, USA}

\author[0000-0003-3460-0103]{Alexei V. Filippenko}
\email{}
\affiliation{Department of Astronomy, University of California, Berkeley, CA 94720-3411, USA}

\begin{abstract}

Correlations between standardized Type Ia supernova (SN Ia) luminosities and host-galaxy properties are routinely modeled to avoid bias in cosmological parameter inference. A recent hypothesis attributes these correlations to progenitor-age variations and, combined with a strong ($\sim$5–6~Gyr) age evolution between low- and high-redshift samples, could alter cosmological conclusions. 
We test this scenario using the SN Ia host galaxies of TITAN DR1, the largest low-redshift sample of its kind to date (6,983 hosts; $0\lesssim z \lesssim 0.15$). Progenitor ages are estimated by combining host-galaxy star-formation histories (SFHs) with empirical delay-time distributions. The SFHs are constrained via spectral energy distribution (SED) fitting of photometry spanning ultraviolet (UV) to mid-infrared (MIR) wavelengths, enabling robust separation of dusty star-forming and quiescent systems. 
The resulting progenitor-age distribution has a mean of 3.5~Gyr, substantially younger than predicted by strong-evolution models. It is strongly peaked near 2.2~Gyr, predominantly from star-forming hosts (60\% of the sample), with a smaller, broader component centered near 6.0~Gyr from quiescent systems. Restricting to high-mass galaxies (in order to isolate progenitor effects from the mass-step), the age difference between host types reduces to 3.3~Gyr which, under the age-dependence hypothesis, would imply a 0.10~mag luminosity offset, inconsistent with observed standardized magnitudes. We infer a modest 1.5~Gyr evolution in mean progenitor age over cosmic time which, combined with observed age–Hubble-residual (HR) relations, yields a maximum redshift-dependent bias of $\Delta\mathrm{HR}=-0.007^{+0.012}_{-0.014}$~mag, consistent with zero. We find no evidence for a large, unmodeled progenitor-age systematic beyond what is already captured, to good approximation, by standard host-mass corrections.

\end{abstract}

\keywords{Distance indicators (394), Standard candles (1563), Photometry (1234), Spectral energy distribution (2129)}

\section{Introduction} \label{sec:intro}

Type Ia supernovae (SNe~Ia), energetic thermonuclear explosions of white dwarfs (WDs), are among the most successful standardizable candles for cosmology \citep[see][for reviews]{Filippenko_2005_SNIacosmology, Howell_2011_NatureReview, Branch_Wheeler_2017_Supernova}. 
Since their early success in probing the expansion history of the universe \citep{Riess_1998_DE,Perlmutter_1999_DE}, decades of effort by the community have improved the standardization framework. The most commonly employed method uses the color (e.g., $c$) and stretch (e.g., $x_1$) of each SN, measured by fitting a light-curve model \citep[e.g., SALT;][]{Guy_2007_SALT2,Kenworthy_2021_SALT3} to observed time-series of multiband photometry. Correcting for the empirical color--luminosity \citep{Riess_1996_MLCSs} and stretch--luminosity \citep{Phillips1993} relations with linear coefficients $\alpha$ and $\beta$ \citep{Tripp1998},
\begin{equation}\label{eq:Tripp}
    m_{B,\text{Tripp}} = m_B + \alpha x_1 - \beta c\, ,
\end{equation}
reduces the spread (scatter) of the inferred luminosity from $\sim0.5$--0.8~mag to $\sim0.15$\ mag. The population-level observational biases, $\Delta\mu_\mathrm{bias}$, owing to target selections, observing conditions, selection cuts, and classification efficiencies, are quantified via forward-modeling \citep[e.g.,][]{Kessler2009,Kessler_Scolnic_2017_BBC} and are accounted for in the distance moduli,
\begin{equation} \label{eq:distmod}
    \mu_\mathrm{SN} = m_{B,\text{Tripp}} - M_0 + \Delta\mu_\mathrm{bias}\, ,
\end{equation}
where $M_0$ is the $B$-band absolute magnitude when $c=x_1=0$.
The consistency and stability of the residuals from the predictions of cosmological models (Hubble residuals; $\mathrm{HR} \equiv \mu_\mathrm{SN}-\mu_\mathrm{cosmo}(z)$) across different redshift ranges is therefore crucial in conducting reliable cosmological experiments, and any residual trend that could evolve over redshift requires a thorough investigation to quantify the size (or lack thereof) of systematic biases.

One of the most commonly discussed sources of such redshift-dependent residual trends in HR is the empirical relationship between HRs and the host galaxies' physical properties. 
SNe~Ia occurring in massive hosts are found to be more luminous by $\sim0.05$--0.10 mag \citep{Sullivan2010,Kelly2010,Lampeitl2010} after standardization, with the dividing line usually drawn at $\log_{10}(M_*/M_\odot)=10$; this phenomenon is often referred to as the ``mass-step.'' The trend is widely interpreted as an empirical proxy for more physical causes, as HRs have been reported to correlate with several other host-galaxy properties, including specific star-formation rates (sSFRs; e.g., \citealt{Rigault2020}), metallicity \citep[e.g.,][]{childress_2013_massstep, MorenoRaya2016}, galactocentric distance \citep[e.g.,][]{Toy2025}, rest-frame colors \citep[e.g.,][]{Kelsey2021, Kelsey2023, Ginolin2025_colour}, morphology \citep[e.g.,][]{Murakami_2021_SNIa_host}, spectroscopic features \citep[e.g.,][]{Martin2024, Dixon2025}, and others.

Large contemporary cosmological analyses nevertheless find that including a mass-based correction removes the bulk of host-dependent HR trends \citep{Betoule2014,Scolnic2018,Brout2019,Popovic2025}. Many of these studies find no additional statistically significant correlations required for cosmological inference at current precision. These analyses likewise find no statistically significant redshift evolution in the mass-step size \citep{Vincenzi2024, Rubin2025, Popovic2025}. 

Mass-based corrections are particularly attractive from a practical standpoint, as host-galaxy stellar mass can be robustly and homogeneously inferred from broad-band photometry across the full redshift range probed by current SN Ia samples \citep[e.g.,][]{Taylor2011_GAMA}, in contrast to other host-galaxy properties (e.g., SFR, stellar population age, dust extinction $A_V$, or morphology) that are more difficult to measure reliably at higher redshift owing to the limited wavelength coverage, signal-to-noise ratio (S/N), and spatial resolution of available photometry.

As the sample size of SNe~Ia increases, however, uncertainties around the physical cause of host dependency and the exact form of its redshift evolution have become a limiting factor in cosmology. One of the most recent and successful host-dependency models attributes this mass dependency to differences in interstellar medium (ISM) extinction laws \citep{Brout_Scolnic_2021,Popovic_2021_hostBBC}. While successful at removing the majority of host- and $c$-dependent trends in HR, studies have shown the presence of a mass-step in the near-infrared (NIR) which cannot be fully explained by the dust model alone \citep[e.g.,][]{Jones_2022_RAISIN}.
The most contemporary cosmological analysis of SNe~Ia \citep{Popovic2025} finds SN~Ia astrophysics, including the uncertainty on host dependency, to be the dominant source of systematic uncertainty in the measurement of $w$, the equation-of-state parameter for dark energy.
With Stage IV dark-energy surveys such as the Legacy Survey of Space and Time (LSST; \citealt{Ivezic2019_LSST}) and the {\it Euclid} \citep{Laureijs2011_Euclid} and {\it Roman} \citep{Spergel2015_WFIRST, Akeson2019_WFIRST} space telescopes, cosmological constraints will become increasingly limited by astrophysical and methodological systematics rather than statistical precision --- a potential drawback in the era of cosmological tensions \citep[e.g.,][]{Riess_2022_SH0ESmain,Murakami_2023_SIP,Breuval_2024_SMC,DES_2024_SNIa_Y5,DESI_BAO}.

In search for improved models, recent findings suggest that host dependency is at least partially tied to the intrinsic difference in the progenitor population. For example, \citet{Larison2024} report that SNe Ia hosted in older, quiescent galaxies in the centers of nearby, rich galaxy clusters constitute an intrinsically different population from field SNe Ia, with an almost exclusively low-$x_1$ distribution. These low-$x_1$ objects' stretch-luminosity law has been shown to possibly differ from the high-$x_1$ counterparts \citep{Garnavich2023, Larison2024, Ginolin2025_stretch,Murakami_Scolnic_2025_ZTF}; such residuals could be absorbed (at least in part) into the empirical host-mass correction. Spectroscopic features of SNe~Ia \citep{Siebert2020, Dettman2021, Murakami_2023_SIP} and of host galaxies \citep{ Martin2024,Dixon2025} have been found to explain a fraction of HRs after applying the mass-step correction which further indicates ties to the progenitor physics.

One physically plausible explanation to such intrinsic effects is the age of the progenitor system. The supernova progenitor age distribution (SPAD) can be estimated by convolving a galaxy’s star formation history (SFH) - which describes the rate at which a galaxy forms stars over cosmic time - with the delay-time distribution (DTD), which gives the probability distribution of time delays between star formation and the resulting supernova event. A series of studies on the host galaxies' age \citep[e.g.,][]{Rose_2019_age,Kang_2020,Rose_2020_age_rebuttal,Lee_2020,Rigault_2020_SFRdependency,Murakami_2021_SNIa_host,Zhang_2021_ageslope,Lee2022,Briday2022,Wiseman2022,Wiseman2023,Chung2025,Son2025} have discussed the effect of population ages on HRs, the ties to progenitor physics, and implications for cosmology.
\citet[hereafter \Chung]{Chung2025} find that HRs (prior to mass-step and bias corrections) correlate strongly with host stellar population age, and argue that progenitor/host age is a primary driver of correlations between HRs and host-galaxy properties. Using this HR--age relation, \citet[hereafter \Son]{Son2025} explicitly propagate an age-based correction into cosmological inference by modeling the evolution of progenitor ages, claiming a $\sim0.16$~mag bias at $z\approx 1$ and a significant change in the fitted cosmology.

The methods presented by \Son have been directly challenged;
\citet[hereafter \Wiseman]{Wiseman2026} present a detailed reanalysis, finding that
(i) applying mass-step and bias corrections removes most of the age dependency, resulting in a statistically insignificant slope of $\Delta\mathrm{HR}\ [\mathrm{mag}]=-0.007^{+0.012}_{-0.014}\cdot\Delta t_\mathrm{MWA}\ [\mathrm{Gyr}]$, where $t_\mathrm{MWA}$ is the mass-weighted age of the host galaxy, and
(ii) simulations show that host-galaxy stellar ages differ significantly from SN Ia progenitor ages. Consequently, an HR--age slope measured with respect to $t_{\rm MWA}$ cannot be directly interpreted as a progenitor-age--luminosity relation.
Based on these findings, \Wiseman\ conclude that the large redshift-dependent luminosity corrections proposed by \Son\ are not supported by the data. Instead, existing measurements indicate that SN~Ia cosmology remains robust to progenitor-age evolution at the level relevant for current dark-energy constraints, and that a physically consistent mapping between progenitor properties and host-galaxy observables is required before invoking age-dependent cosmological corrections.

The \textit{Type Ia Supernova Trove from ATLAS in the Nearby Universe} \citep[TITAN; Murakami et al., in prep.; Tweddle et al., in prep.;][]{Marlin2025} dataset is a new, low-$z$ ($0\lesssim z\lesssim0.15$) SN Ia sample that enables a novel study to shed light on this topic through a data-driven, empirical approach. TITAN contains 8,378 spectroscopically confirmed SNe~Ia, the largest sample of its kind. The majority of the SNe are associated with a host galaxy through a robust, systematic method (Tweddle et al., in prep.), the properties of which --- when combined with the large sample size and narrow redshift range --- provide an ideal opportunity to study the full range of potential SN Ia progenitor ages.

In this work, we measure SFHs of 6,983 TITAN hosts and infer progenitor ages using an SN~Ia DTD from the literature.
Sec.~\ref{sec:SFH} discusses our multiwavelength dataset for TITAN hosts, the resulting spectral energy distributions (SEDs), and SED modeling methods used to determine SFHs. The review and choice of our baseline DTD is discussed in Sec.~\ref{sec:DTD}. We present the inferred progenitor-age distributions and direct tests of age-based HR predictions in Sec.~\ref{sec:age}.
Sec.~\ref{sec:discussion} highlights key considerations required to avoid biased progenitor ages and to enable a physically consistent comparison between our measurements and model predictions. In Sec.~\ref{sec:cosmo}, we explore simple models to reassess the possible systematic effects of progenitor-age evolution in cosmology, and emphasize the importance of a full, data-driven, forward-modeling approach to properly account for observational biases and physical evolution simultaneously. We summarize our main conclusions in Sec.~\ref{sec:conclusion}.

\section{Star-formation histories of TITAN host galaxies} \label{sec:SFH}

\begin{figure*}
    \centering
    \includegraphics[width=1\linewidth]{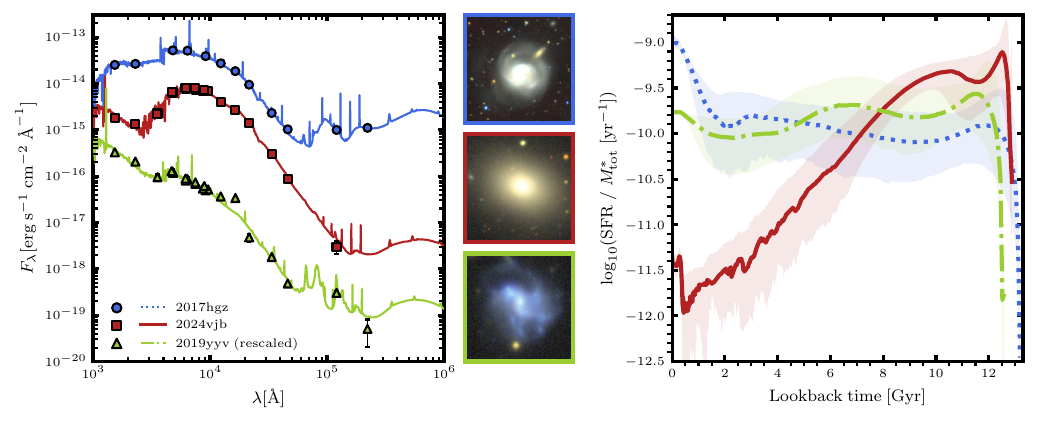}
    \caption{Example best-fit SEDs and SFHs for the late-type host galaxy of SN~2017hgz, the early-type host galaxy of SN~2024vjb, and the irregular host galaxy of SN~2019yyv with clumpy star-forming regions, consistent with a disturbed system and suggestive of a possible merger. \textit{Left:} Observed FUV to MIR photometry of the three SN~Ia host galaxies (black-edged points), shown with the median best-fit SED models (blue for SN~2017hgz, red for SN~2024vjb, green for SN~2019yyv). To better facilitate a direct comparison of SED shapes, the fluxes for the host galaxy of SN~2019yyv have been reduced by a factor of 5. \textit{Middle:} Cutout images of the three host galaxies to illustrate their distinct morphologies. Colored frames correspond to the different curves in the left and right panels. \textit{Right:} Corresponding SFHs, shown as the SFR normalized by the total stellar mass of each galaxy as a function of lookback time.}
    \label{fig:sed_to_sfh}
\end{figure*}

The TITAN sample and its validation as a low-redshift SN~Ia cosmology dataset ($0 \lesssim z \lesssim 0.15$) are introduced by \citet{Marlin2025}, who demonstrate that the photometric calibration and light-curve quality meet the precision requirements for cosmological analyses. The TITAN sample comprises 8,378 spectroscopically confirmed SNe Ia with high-cadence, dual-band light curves obtained by the ATLAS all-sky survey \citep{TonryATLAS2018,SmithATLAS2020}. Dedicated papers describing the full catalog, light-curve data release, host-galaxy matching and analysis framework, and survey simulations are forthcoming (Murakami et al., in prep.; Tweddle et al., in prep.).

In this paper, we make use of the extensive TITAN host-galaxy catalog (Tweddle et al., in prep.). For each event, the most probable host is identified using a combination of the directional light radius (DLR) method \citep{Sullivan2010, Gupta2016} and redshift consistency between the SN and candidate galaxies assembled from multiple catalogs. For each host, homogeneous archival broad-band photometry spanning the far-UV and near-UV (\textit{GALEX} FUV + NUV; $\lambda_{\mathrm{min}} = 0.153~\mu\mathrm{m}$) to the MIR (\textit{WISE} W1--W4; $\lambda_{\mathrm{max}} = 24~\mu\mathrm{m}$) is then compiled using \texttt{HostPhot} \citep{MullerBravo2022}.

The resulting SEDs are modeled using the Bayesian Analysis of Galaxies for Physical Inference and Parameter EStimation (\texttt{BAGPIPES}; \citealt{Bagpipes}) framework. In Tweddle et al. (in prep.), a nonparametric SFH formalism with a continuity prior \citep{Leja2019} is adopted, which regularizes the change in $\log_{10}(\mathrm{SFR})$ between adjacent time bins. However, the discretized (binned) nature of this SFH parameterization imposes a finite temporal resolution on the inferred SFH. When multiplied by a DTD to derive the supernova progenitor age distribution (SPAD), this discretization propagates directly into the resulting age distribution where the sharp edges between adjacent SFH bins are effectively mapped into abrupt changes in the SPAD, producing a characteristic sawtooth-like appearance rather than a smooth distribution. This binning-induced structure is not physical, but rather a consequence of the limited temporal resolution of the SFH model. It could bias summary statistics such as the median progenitor age, which become sensitive to the arbitrary placement and width of the SFH bins.

Therefore, in this work, we instead adopt an \cite{Iyer_2019_nonparamSFH} SFH formalism, which provides a smoother and more continuous representation of the SFH using a Gaussian process in $\log_{10}(\mathrm{SFR})$ as a function of lookback time. This model treats the SFH as a stochastic function drawn from a covariance kernel that correlates star formation at nearby times, with hyperparameters that control the overall smoothness and characteristic timescale of variation. Therefore, the model allows for the flexible, nonparametric reconstruction of complex SFHs without imposing discrete time bins, yielding a correspondingly well-behaved SPAD that is less sensitive to discretization effects. We show, in Appendix \ref{sec:app_iyer}, that this alternative choice of SFH does not materially affect the inferred properties of TITAN host galaxies compared to those derived by Tweddle et al. (in prep.), nor key statistical quantities associated with the progenitor age distributions derived in the remainder of this work.

Of the full TITAN sample, a total of 7,291 (87.0\%) TITAN SNe Ia possess a confident host-galaxy association whose SED is successfully fitted, of which 6,983 (83.3\%) possess sufficient multiwavelength photometric coverage (UV--MIR) for reliable age estimates to be obtained. These galaxies constitute the final sample analyzed in this work. A detailed description of the full host SED modeling procedure, parameter priors, and validation tests is provided by Tweddle et al. (in prep.).

This subsample of TITAN is well suited to the analysis of SN Ia progenitor ages because all hosts have broad UV--MIR photometric coverage, enabling more reliable SED-based SFH inference than is possible with optical-only data. As demonstrated for SN Ia hosts at low and high redshift \citep[Tweddle et al., in prep.;][]{Ramaiya2025}, inferred SFRs and dust attenuation can be strongly biased when the dust-emission regime is not constrained. In the TITAN redshift range ($z \lesssim 0.15$) MIR measurements, which directly constrain dust-reprocessed starlight, are therefore crucial and reduce the most severe age-dust degeneracies in broad-band SED fitting \citep[e.g.,][]{Papovich2001, Burgarella2005}. High-S/N spectroscopy remains the gold standard for further disentangling age, metallicity, and dust via absorption- and emission-line diagnostics, such as Balmer lines and the 4000\,\AA\ break \citep[e.g.,][]{Maraston2020_sdss, Sanchez2016_pipe3d}, but homogeneous integral-field-unit spectroscopy is typically unavailable for complete SN~Ia host samples used in modern cosmology.

For progenitor-age inference, the dominant practical failure mode is the misclassification of dusty star-forming galaxies as quiescent (or vice versa), which can strongly distort inferred recent SFHs and therefore delay-time-weighted progenitor-age estimates. MIR/far-IR (FIR) constraints provide an orthogonal handle on obscured star formation and materially improve the robustness of the inferred SFHs and derived progenitor ages relative to optical/UV-only SED fitting \citep[e.g.,][]{Conroy2013, Leja2017, daCunha2008, Ciesla2015} while remaining practical to apply homogeneously to large samples. We quantify the size of these effects for TITAN in Sec.~\ref{sec:need_mir}.

Figure \ref{fig:sed_to_sfh} shows representative best-fit SEDs and the corresponding inferred SFHs for three TITAN SN~Ia hosts spanning the diversity of the sample: (i) a late-type, actively star-forming galaxy, (ii) an early-type, quiescent system, and (iii) an irregular host. In all three cases, the fitted SEDs reproduce the observed FUV--MIR photometry, simultaneously constraining the UV light from young stellar populations and the dust-reprocessed emission at longer wavelengths. The inferred SFHs are correspondingly distinct: the late-type host exhibits strong recent star formation, the quiescent host is consistent with predominantly early assembly followed by long-term suppression of star formation, and the irregular host displays a quasiperiodic, bursty SFH suggestive of a more stochastic or interaction-driven evolutionary history. The derived physical parameters (Table~\ref{tab:example_hosts}) align with these qualitative trends, providing a sanity check that our modeling setup yields plausible SFHs across the range of TITAN host morphologies.

\begin{deluxetable}{lccc}
\tablecaption{Representative physical properties inferred from SED fitting for the hosts shown in Figs. \ref{fig:sed_to_sfh} and \ref{fig:sfh_to_age_demo}. Values correspond to the 50$^\mathrm{th}$ percentile of the posterior distributions, and uncertainties represent the 16$^\mathrm{th}$ and 84$^\mathrm{th}$ percentiles, respectively. A DTD with $t_\mathrm{min}=40\ \mathrm{Myr},\ \beta_\mathrm{DTD}=-1.07$ is assumed when computing $\mathbb{E}[t_\mathrm{prog}]$.}
\label{tab:example_hosts}
\tablehead{
\colhead{Property} & \colhead{SN~2017hgz} & \colhead{SN~2024vjb} & \colhead{SN~2019yyv}
}
\startdata
$\log_{10}\!\left(\frac{M_\star}{M_\odot}\right)$ & $10.61^{+0.06}_{-0.06}$ & $11.17^{+0.01}_{-0.02}$ & $9.83^{+0.05}_{-0.05}$ \\
$\log_{10}\!\left(\frac{\mathrm{sSFR}}{\mathrm{yr^{-1}}}\right)$ & $-9.01^{+0.11}_{-0.09}$ & $-11.42^{+0.11}_{-0.10}$ & $-9.77^{+0.06}_{-0.05}$ \\
$A_V\ \mathrm{[mag]}$ & $0.91^{+0.08}_{-0.07}$ & $0.06^{+0.03}_{-0.04}$ & $0.04^{+0.01}_{-0.01}$ \\
$g-z\ \mathrm{[mag]}$ & $0.90^{+0.03}_{-0.04}$ & $1.18^{+0.01}_{-0.02}$ & $0.40^{+0.02}_{-0.03}$ \\
$t_{\mathrm{MWA}}\ \mathrm{[Gyr]}$ & $4.60^{+0.99}_{-0.92}$ & $10.43^{+0.52}_{-0.75}$ & $7.15^{+0.93}_{-1.01}$ \\
$\mathbb{E}[t_\mathrm{prog}]\ \mathrm{[Gyr]}$ & $0.49^{+0.20}_{-0.14}$ & $9.59^{+0.64}_{-0.76}$ & $2.12^{+0.55}_{-0.55}$ \\
\enddata
\end{deluxetable}

\section{The SN Ia Delay-Time Distribution} \label{sec:DTD}

A central ingredient in our inference is the SN Ia DTD, which specifies the relative probability that a stellar population produces an SN Ia after a delay time $\tau$. The DTD encapsulates the combined effects of stellar evolution, binary-interaction physics, and explosion mechanisms, and therefore provides a powerful, indirect probe of SN Ia progenitor channels \citep{Maoz2014_DTDreview}. In this work, the DTD enters only as a weighting function that maps each host galaxy's inferred SFH into a posterior over progenitor ages (Sec.~\ref{sec:age}); we therefore require a DTD that captures the empirically established preference for shorter delays, rather than an absolute rate normalization.

Observationally, the DTD is commonly constrained through measurements of the redshift evolution of the volumetric SN Ia rate and deconvolving with the cosmic SFH \citep[CSFH; e.g.,][]{GalYam2004, Perrett2012, Frohmaier2019} through the relation
\begin{equation}
R_{\mathrm{Ia}}(t) = \int_0^t \Phi(t - \tau)\Psi(\tau) \mathrm{d}\tau \label{eq:rate}
\end{equation}
where $R_{\mathrm{Ia}}(t)$ is the volumetric SN Ia rate at cosmic time $t$, $\tau$ is the delay time, $\Phi$ is the CSFH, and $\Psi$ is the DTD. This equation applies not only to a cosmological volume, but also to individual galaxies; in this instance the CSFH is replaced with the SFH of individual galaxies \citep{Brandt2010, Maoz2011, Wiseman2021}. 

In practice, the DTD is modeled as a truncated power law,
\begin{equation}\label{eq:DTD}
    \Psi(t) = \left\{
\begin{array}{ll}
      0 & t \leq t_\mathrm{min} \\
      \alpha_\mathrm{DTD} (\frac{t}{\mathrm{Gyr}})^{\beta_\mathrm{DTD}} & t > t_\mathrm{min}\, , \\
\end{array} 
\right.
\end{equation}
where $\beta_\mathrm{DTD}$ is the slope of the DTD and $t_{\mathrm{min}}$ represents the minimum delay time required to produce an SN Ia explosion from the formation of the initial stellar population.
Stellar evolution models typically place this minimum delay at $t_{\mathrm{min}} \approx 40~\mathrm{Myr}$ for single-degenerate channels \citep[e.g.,][]{Wang2012DTD, Ruiter2009DTD, Bours2013DTD, Mennekens2010DTD, Yungelson2010DTD}. Several of these models also predict similar minimum delays for double-degenerate channels, although the detailed value depends on assumptions regarding binary evolution and common-envelope physics. Observational studies consistently find values for $t_\mathrm{min}$ in the range 40--200 Myr \citep[e.g.,][]{Aubourg2008, Castrillo2021, Chen2021_delaytime, Wiseman2021}. The normalization $\alpha$ cancels in the per-galaxy progenitor-age probabilities used in this work (see Sec.~\ref{sec:age}) and is therefore not important for our main results.

Empirical measurements consistently find an approximately $t^{-1}$ behavior over $\sim0.1$--10 Gyr across a wide range of environments \citep[e.g.,][]{Totani2008, Maoz2010, Maoz2012, Maoz2014_DTDreview}. Unless explicitly stated otherwise, throughout this work we adopt a fiducial slope $\beta_\mathrm{DTD} = -1.07$ \citep{Maoz2012} and $t_\mathrm{min}=40\ \mathrm{Myr}$, and justify this choice of $t_\mathrm{min}$ for our baseline result in Appendix \ref{sec:app_dtd}. The effects of alternative DTD parameter choices are discussed in detail in the following section.

\begin{figure}
    \centering
    \includegraphics[width=\linewidth]{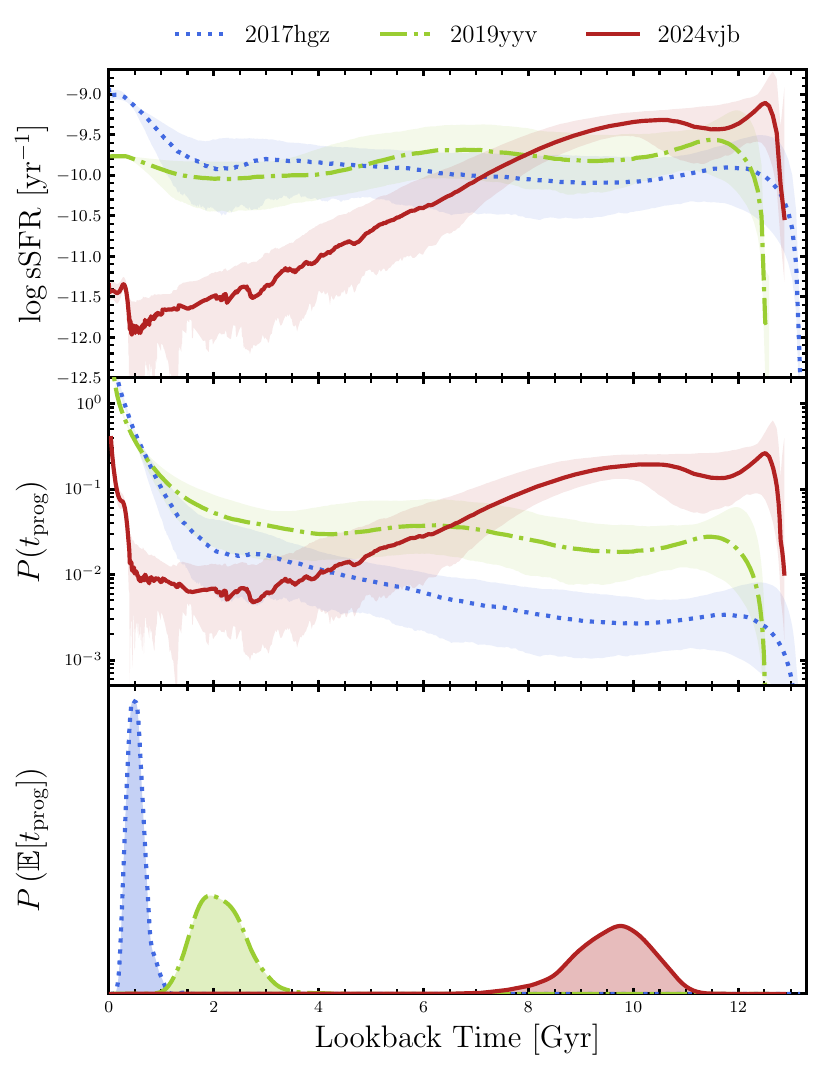}
    \caption{Illustration of how an inferred host-galaxy SFH maps to a delay-time-weighted SN Ia progenitor-age probability. \textit{Top:}  SFHs for the three host galaxies in Fig.~\ref{fig:sed_to_sfh}, shown as the SFR per unit present-day stellar mass as a function of lookback time. 
    \textit{Middle:} Inferred SN progenitor-age distribution (SPAD; $P(t_\mathrm{prog})$), for each host, derived by multiplying SFHs in the top panel by the DTD ($t_\mathrm{min}=40\ \mathrm{Myr},\ \beta_\mathrm{DTD}=-1.07$).
    \textit{Bottom:} Probability density of $\mathbb{E}[t_\mathrm{prog}]$, the expected mean progenitor age from each host. This distribution is inferred by sampling 100 SNe each from 1000 SPAD chains for each galaxy, encapsulating the uncertainty in the SFHs.}
    \label{fig:sfh_to_age_demo}
\end{figure}

\section{Measurements of Progenitor Ages of TITAN SNe Ia} \label{sec:age}

\begin{figure*}
    \centering
    \includegraphics[width=\linewidth]{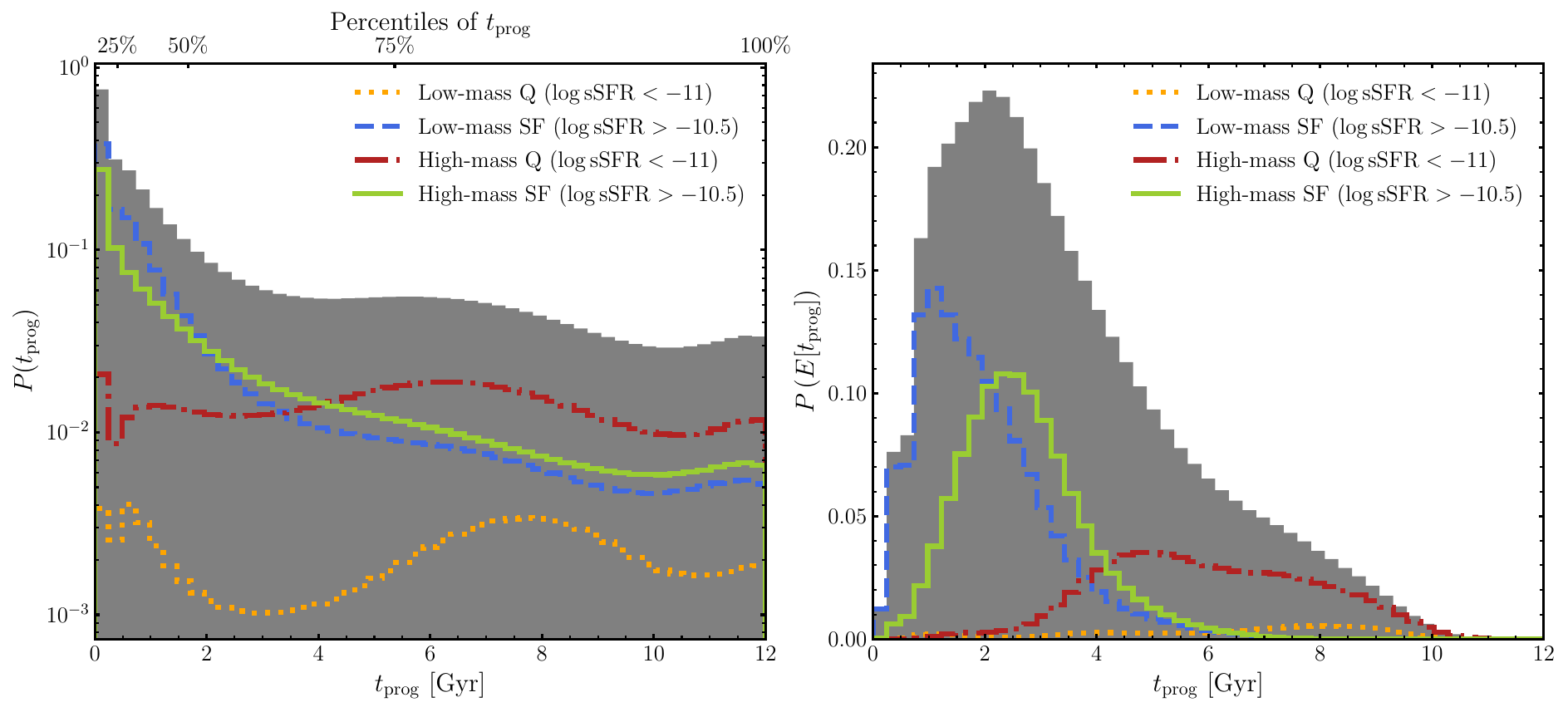}
    \caption{Distributions of SN~Ia progenitor ages in the TITAN sample, inferred by multiplying host-galaxy SFHs by a DTD characterized by $t_\mathrm{min}=40\ \mathrm{Myr}$ and $\beta_\mathrm{DTD}=-1.07$. Gray histograms show results for the full sample, while colored curves indicate subsamples split by stellar mass and sSFR (with low- and high-mass galaxies defined by $\log(M_*/M_\odot)<10$ and $>10$, and star-forming and quiescent systems defined by $\log(\mathrm{sSFR})>-10.5$ and $\leq -11$, respectively). As in Fig.~\ref{fig:sfh_to_age_demo}, each host contributes 100 realizations drawn from each of 1000 \texttt{Bagpipes} posterior SFHs, ensuring that uncertainties in the SFH reconstruction are fully propagated. \textit{Left:} The stacked supernova progenitor-age distribution (SPAD), representing the probability distribution from which the age of any individual SN~Ia in the sample is drawn. The distribution has a median of 1.9~Gyr and exhibits clear bimodality, reflecting the distinct contributions of young progenitors in star-forming galaxies and older progenitors in quiescent systems. \textit{Right:} The stacked distribution of the expected progenitor age, $P(\mathbb{E}[t_\text{prog}])$. The mean of this distribution is 3.5~Gyr, corresponding to the average progenitor age across the TITAN sample. Star-forming galaxies preferentially host younger progenitors, with mean age 2.2~Gyr, while quiescent galaxies host systematically older progenitors, with mean age 6.0~Gyr. These trends persist across stellar-mass bins.}
    \label{fig:age_histogram}
\end{figure*}

For each TITAN host galaxy, we compute a probability density describing the progenitor age of the observed TITAN SN Ia. The quantity of interest is not the SN~Ia production rate of the host as a function of lookback time, but rather the probability that the observed SN~Ia originates from a stellar population of age $t_\mathrm{prog}$, as well as the average age of progenitors from across the host population. The probability is equivalent to the integrand of Eq. \ref{eq:rate}, normalized over all allowed progenitor ages:
\begin{equation}
P_i\left(t_\mathrm{prog} \mid \mathrm{SN}\right)
= \frac{\Phi(t_\mathrm{prog})\Psi_i(t_\mathrm{prog})}
{\int_0^{t_\mathrm{univ}(z_\mathrm{gal})} \Phi(t)\Psi_i(t)\mathrm{d}t} \, ,\label{eq:age_prob}
\end{equation}
where $\Psi$ and $\Phi$ are defined in terms of lookback time, and $t_\mathrm{univ}(z_\mathrm{gal})$ is the age of the universe at the redshift of the host galaxy. We evaluate this expression on a fine, equally spaced grid in $t_\mathrm{prog}$ by resampling the best-fitting continuity SFH. 

For subsequent analyses, we consider two closely related but distinct summary statistics: (i) the finite-mixture distribution of progenitor ages $P(t_\text{prog}\mid\mathrm{SN})$, and (ii) the ensemble expectation value $\mathbb{E}[t_\mathrm{prog}]$. These quantities capture complementary statistical information and are suited to different types of analyses.

The mixture distribution $P(t_\mathrm{prog})$ is constructed by linearly combining the individual progenitor-age probability density functions (PDFs) (Eq.~\ref{eq:age_prob}) from across the sample,
\begin{equation} \label{eq:mixture_pdf}
    P(t_\text{prog}|\text{SNe}) = \frac{1}{N_\mathrm{SN}}\sum P_i\left(t_\mathrm{prog} \mid \mathrm{SN}\right),
\end{equation}
where $N_\mathrm{SN}$ is the total number of SNe in the sample. The weights are kept the same for all hosts because the TITAN dataset already reflects the selection effects owing to the overall SN~Ia rate per host. This distribution therefore represents the progenitor-age probability density for a randomly selected SN~Ia in the sample and, in the large-sample limit, in the local universe.

In parallel, we compute an unbiased estimator of the progenitor age for each system via the expectation value of its individual PDF,
\begin{equation} \label{eq:age_ev}
    \mathbb{E}[t_{\text{prog},i}] = \int_0^{t_\text{univ}} P(t_\mathrm{prog})_i \ t_\mathrm{prog} \ \mathrm dt_\mathrm{prog}\, .
\end{equation}
For both quantities, we propagate SFH uncertainties by bootstrapping the computation of Eqs. \ref{eq:age_prob} and \ref{eq:age_ev} over parameter chains from \texttt{Bagpipes}. This procedure yields a combined (mixture) PDF of $P(t_\mathrm{prog})$ as well as a sample of $\mathbb{E}[t_\mathrm{prog}]$ values, from which we produce a posterior probability distribution of expectation values, $P\left(\mathbb{E}[t_\mathrm{prog}]\right)$. 
This quantity provides two distinct and important insights into the properties of $t_\text{prog}$: its galaxy-to-galaxy difference highlights the dependency of $t_\text{prog}$ on galaxy properties, and when stacked over multiple hosts, it visualizes the similarity within or differences across subsamples. The mean $t_\text{prog}$ of this distribution is identical to the mean of all SNe in TITAN samples.

\begin{deluxetable*}{cc c c  c c}  \label{tab:DTD_variants_stats}
    \setlength{\tabcolsep}{8pt} 
    \tablecaption{Distribution of sampled SN ages, expectation values, and the overall mean progenitor ages for different choices of DTD parameters. The boldfaced row represents our baseline result. All of our variants, including the \Son DTD, have median values $>3$~Gyr younger than the $\sim6.5$~Gyr median age at $z\approx0$ shown by Fig.~2 in \Son.}
    \tablehead{
    \multicolumn{2}{c}{DTD} &
    \colhead{$P(t_\text{prog})$} &
    \colhead{$P(\mathbb{E}[t_\text{prog}])$} & \colhead{$\langle t_\text{prog}\rangle$} & \colhead{Note} \\
    $t_\text{min}$ [Myr] & $\beta$ & 16--50--84$^{\mathrm{th}}$ [Gyr] & 16--50--84$^{\mathrm{th}}$ [Gyr] & [Gyr] &\\
     & & \textit{SN-to-SN variation} & \textit{host-to-host variation} & \textit{snapshot mean at $z\approx0$} &
    }
    \startdata
    40  & $-$1.00 & 0.3, 2.2, 8.0 & 1.5, 3.3, 6.0 & 3.6 & \\
    \bf 40 & \bf $-$1.07 & \bf 0.2, 1.9, 7.7 & \bf 1.4, 3.0, 5.7  &  \bf 3.5 & baseline \\
    40  & $-$1.13 & 0.1, 1.6, 7.5 & 1.3, 2.7, 5.5  & 3.3 &\\
    300  & $-$1.00 & 0.7, 3.2, 8.5 & 2.1, 3.9, 6.6 & 4.3 & \Son DTD\\
    300 & $-$1.07 & 0.7, 3.0, 8.3 & 1.9, 3.7, 6.5 & 4.1 \\
    300 & $-$1.13 & 0.6, 2.7, 8.1 & 1.8, 3.6, 6.4 & 4.0  \\
    \hline
    \multicolumn{2}{c}{\Son at $z\approx0$} & $\sim6.5$ & \nodata & \nodata & \Son Figure~2
    \enddata
\end{deluxetable*}

Figure~\ref{fig:sfh_to_age_demo} illustrates the mapping from SFH to progenitor-age probability and the expectation value PDFs for the three SNe Ia and their host galaxies presented in Figure~\ref{fig:sed_to_sfh}. The middle panel shows the probability that the observed SN originates from a population with age $t_\mathrm{prog}$, and the bottom panel displays the posterior PDF of the expected mean age $\mathbb{E}[t_\mathrm{prog}]$ from each host if we were to observe SNe from this host many times (useful when grouped by similar hosts). The SN~2019yyv example highlights why host-mass-weighted age and progenitor age are not interchangeable: although the host has $t_\mathrm{MWA} = 7.15^{+0.93}_{-1.01}\ \mathrm{Gyr}$, the inferred progenitor age is $\mathbb{E}[t_\mathrm{prog}] = 2.12^{+0.55}_{-0.55}\ \mathrm{Gyr}$, because even modest recent star formation is strongly upweighted by the DTD (see Sec.~\ref{sec:age_vs_age} for further discussion).

In the following section, we examine the distribution of SN~Ia progenitor ages with multiple statistics: percentiles of SN~Ia samples drawn from $P(t_\text{prog})$ that represent distributions of individual SNe, percentiles of the expectation-value distributions $P(\mathbb{E}[t_\text{prog}])$ that represent variations of expected mean $t_\mathrm{prog}$ between hosts, and the overall mean of SN~Ia progenitor ages, $\langle t_\text{prog}\rangle = \mathbb{E}[P(t_\text{prog})] = \frac{1}{N_\text{SN}}\sum_i \mathbb{E}_i[t_\text{prog}]$. Of these quantities, the mean age $\langle t_\text{prog}\rangle$ is most relevant to the size of overall redshift-dependent evolution, and distributions of per-host expectation values $P_{\text{sub}}(\mathbb{E}[t_\text{prog}])$ provide insights into host-dependent differences in $t_\text{prog}$ within and across subsamples split by similar host-galaxy types.

\subsection{Old Universe, Young SNe Ia} \label{sec:old_univ_young_sneia}

Figure~\ref{fig:age_histogram} shows the SPAD, $P(t_\mathrm{prog})$ (left), and distribution of expected mean progenitor ages $P(\mathbb{E}[t_\mathrm{prog}])$ (right), for the full TITAN sample. The distribution of expectation values is strongly weighted toward young progenitors, with a prominent peak at $\sim2\,\mathrm{Gyr}$. This indicates that most SN Ia progenitors in the local Universe are significantly younger than the present cosmic age.

To better interpret this distribution, we also split the sample by conventional host-galaxy properties: $\log_{10}(M_*/M_\odot)=10$ (low/high mass), $\log_{10}(\mathrm{sSFR})>-10.5$ (star-forming; SF), and $\log_{10}(\mathrm{sSFR})<-11.0$ (quiescent; Q). The intermediate range $-11.0\le \log_{10}(\mathrm{sSFR})\le-10.5$ is excluded to minimize contamination from transitional systems (motivated by, e.g., \citealt{Belfiore2018, Davies2019}), which are not included in the SN~Ia samples discussed in Sec.~\ref{sec:age_tests_S25}.

The distribution of expectation values $\mathbb{E}[t_\mathrm{prog}]$ highlights a clear dependence on host type. Star-forming galaxies are tightly clustered around the young peak at $\sim2\ \mathrm{Gyr}$, whereas quiescent galaxies populate an older, broader distribution centered at $\sim6\ \mathrm{Gyr}$. The contrast between SF and Q systems is significantly stronger than any residual dependence on stellar mass within the SF population, indicating that ongoing star formation, rather than total stellar mass, is the primary driver of progenitor age. 

Overall, the $P(\mathbb{E}[t_\mathrm{prog}])$ distribution is highly skewed toward young ages, with a mode of $2.1\ \mathrm{Gyr}$. Averaging over all TITAN hosts and propagating uncertainties via the \texttt{Bagpipes} chains, we infer a mean progenitor age of $3.5\ \mathrm{Gyr}$. This higher mean is driven by a long tail toward older systems, as reflected in the broad host-to-host variation (16$^{\mathrm{th}}$--84$^{\mathrm{th}}$ percentile range in $P(\mathbb{E}[t_\mathrm{prog}])$: 1.4--5.7\ Gyr). Consistent with this skewness, the median of the full progenitor-age distribution $P(t_\mathrm{prog})$ is $1.9\ \mathrm{Gyr}$, underscoring that the SN~Ia population is dominated by young progenitors, with older systems contributing a subdominant but extended tail.

These results are in clear tension with \Chung and \Son, whose model predicts a $z=0$ median progenitor age of $\sim6.5\ \mathrm{Gyr}$. While the precise shape of the SPAD depends on the assumed DTD, the results presented in Table~\ref{tab:DTD_variants_stats} demonstrate that the dominance of young progenitors is not sensitive to reasonable variations in the DTD. Varying $t_\mathrm{min}$ and the DTD slope $\beta_\mathrm{DTD}$ shifts the distribution only modestly, but in all cases the median of $P(t_\mathrm{prog})$ remains $\lesssim 3.2\ \mathrm{Gyr}$. The origin of this discrepancy, along with other differences from the \Son framework, is discussed further in Sec.~\ref{sec:discussion}.

\subsection{Direct Tests of the Age Bias Model}
\label{sec:age_tests_S25}

\begin{figure}
    \centering
    \includegraphics[width=\linewidth]{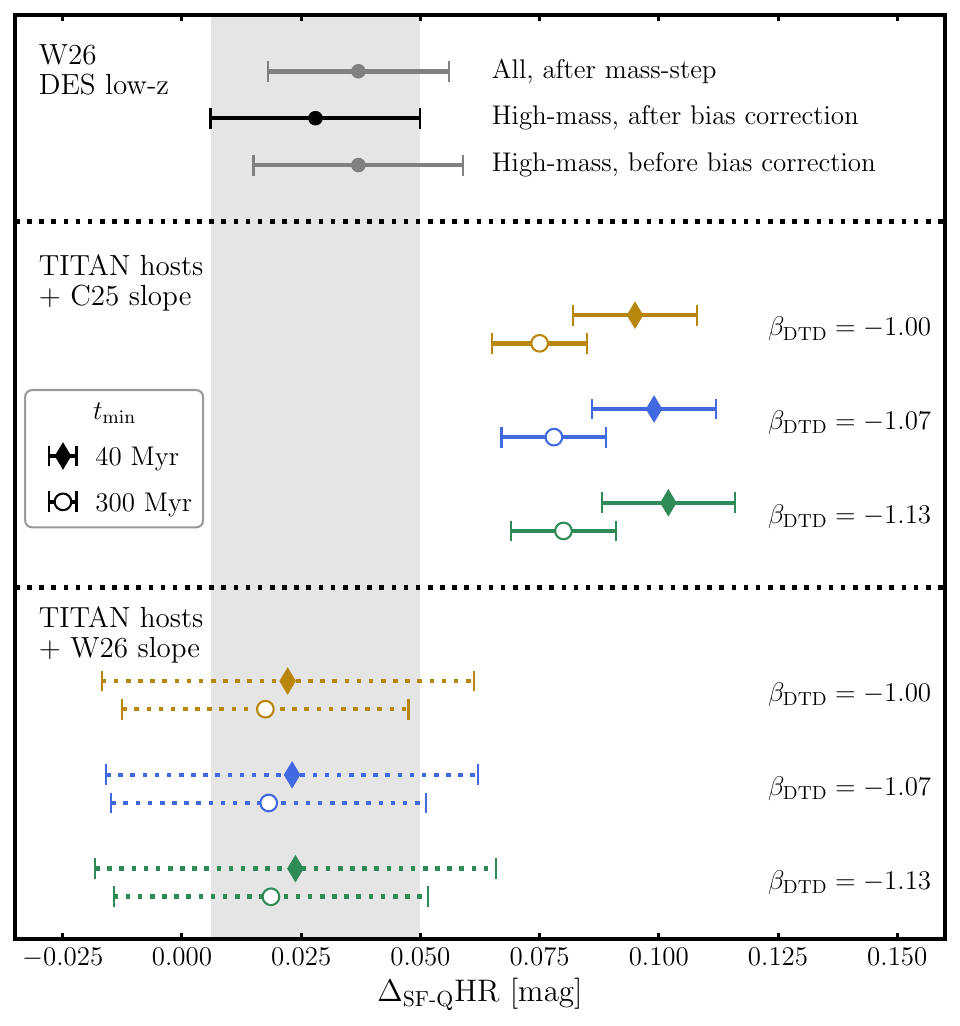}
    \caption{Hubble residual (HR) offsets between SNe~Ia from star-forming and quiescent hosts. \textit{Top}: Direct measurement of DESY5 low-$z$ data \citep{Popovic2025} by \Wiseman. \textit{Middle}: Predictions made by the \Chung and \Son HR--age slope, based on our age measurements. Exact age offsets depend on the choice of DTD, and presented variants cover a wide range of $t_\text{min}$ and $\beta_\text{DTD}$ parameters from the literature. All variants are systematically discrepant with the direct measurement of SN~Ia data (shaded region). \textit{Bottom}: Same as the middle panel, but employing the \Wiseman HR--age slope (measured after applying the mass-step and bias corrections). In both the middle and bottom panels, the age offset refers to the mean offset of the expectation values between high-mass star-forming and high-mass quiescent hosts, as visualized between two subsets in Figure~\ref{fig:age_histogram}.}
    \label{fig:HR_separation_tests}
\end{figure}

Our empirically derived progenitor-age distribution enables a direct test of the magnitude of any progenitor-age-dependent HR offset. In the TITAN sample, when restricting to high-mass hosts, we find a mean progenitor-age difference of $3.30\pm0.03$~Gyr between star-forming and quiescent host populations. This age difference isolates the effect of stellar population age independently of host stellar mass.

\Son\ argue that the observed HR--age relation, measured using host-galaxy mass-weighted stellar age, can be directly interpreted as a progenitor-age--HR relation. Adopting their reported slope,
$\Delta\mathrm{HR}/\Delta t_\mathrm{MWA}= \Delta\mathrm{HR}/\Delta\mathrm{t_\mathrm{prog}}=-0.030\pm{0.004}$~mag~Gyr$^{-1}$,
a $3.30\pm0.03$~Gyr progenitor-age difference would imply an expected HR offset of $0.099 \pm 0.013$~mag between high-mass quiescent and high-mass star-forming hosts.

However, this prediction is not supported by observations. In \Wiseman, using the low-redshift sample employed in the DESY5 cosmological analysis \citep{Popovic2025}, the measured HR offset between morphologically classified star-forming and quiescent hosts is $0.037 \pm 0.019$~mag. Using their data and further restricting to high-mass hosts ($\log_{10}(M_\star/M_\odot)\ge10$), we measure the high-mass-only separation to be $\Delta\mathrm{HR}_{\rm SF-Q} = 0.028 \pm 0.022~\mathrm{mag}$
before bias correction and $\Delta\mathrm{HR}_{\rm SF-Q} = 0.037 \pm 0.022~\mathrm{mag}$ after bias correction. In both cases, the observed separation is $>2.7\sigma$ discrepant with the prediction made by the \Son model. Rather, the observed brightness offset and age difference between the SF and Q populations together favor the \Wiseman slope of $-0.007^{+0.012}_{-0.014}$ mag Gyr$^{-1}$.
In the morphological classification of DES low-$z$ host galaxies in \Wiseman, transitional (e.g., S0) types and unclear cases are excluded from the analysis. Previously discussed cuts for subgroups of the TITAN sample exclude a small fraction ($<20\%$) of TITAN hosts to follow this.

As discussed in Sec.~\ref{sec:old_univ_young_sneia}, the separation of ages between two subsamples (SF and Q) can vary when DTD parameters are changed. In Figure~\ref{fig:HR_separation_tests}, we display the predicted HR separation values for all variants shown in Table~\ref{tab:DTD_variants_stats}, as well as the observed separations measured by \Wiseman. The above discussion is applicable to all of these variants, as they exhibit a systematic offset from the observed HR offsets.

While a definitive assessment of subtle host-type-dependent residuals beyond the standard mass-step correction will benefit from larger, homogeneous samples (e.g., TITAN DR1; Murakami et al., in prep.), this analysis shows that existing low-redshift datasets already disfavor a scenario in which progenitor age is correlated with the Hubble residual at the level of $0.030~\mathrm{mag\ Gyr^{-1}}$. The empirically observed HR offsets between star-forming and quiescent systems are substantially smaller than required by such a model.

\subsection{Progenitor Age $\ne$ Galaxy Age} \label{sec:age_vs_age}
\begin{figure}
    \centering
    \includegraphics[width=\linewidth]{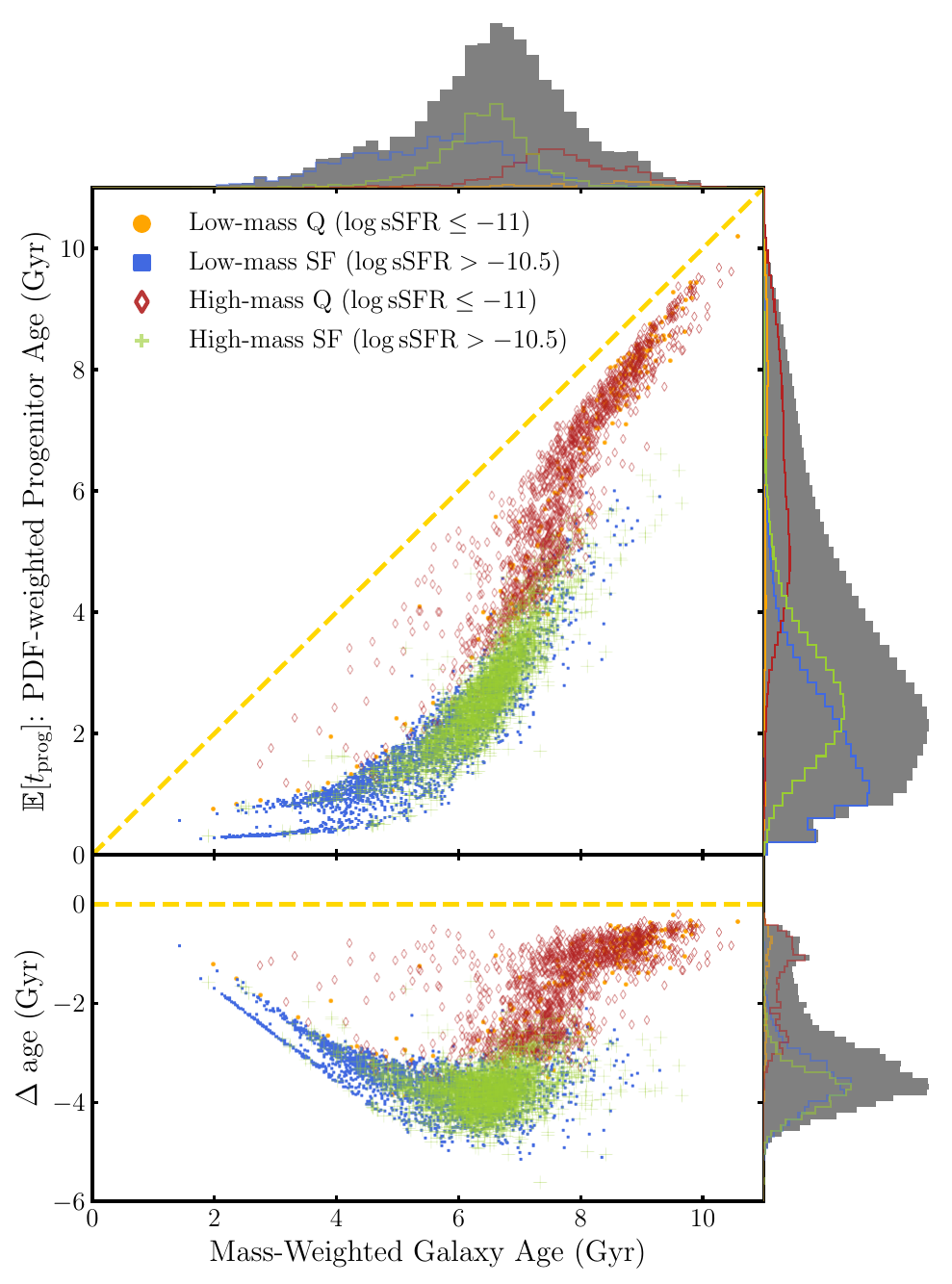}
    \caption{The relationship between SN Ia progenitor age and the mass-weighted age of its host galaxy. Young progenitors $(\mathbb{E}[t_\mathrm{prog}] \lesssim 4\ \mathrm{Gyr})$ are primarily associated with currently star-forming galaxies over a wide range of mass-weighted ages $(1.5 \lesssim t_\mathrm{MWA} \lesssim 8\ \mathrm{Gyr})$, while old progenitors $(\mathbb{E}[t_\mathrm{prog}] \gtrsim 4\ \mathrm{Gyr})$ predominantly originate from quiescent galaxies spanning a narrow range of mass-weighted ages $(8 \lesssim t_\mathrm{MWA} \lesssim 10\ \mathrm{Gyr})$. In both instances, the transformation between progenitor age and mass-weighted age is highly nonlinear, and $\mathbb{E}[t_\mathrm{prog}] < t_\mathrm{MWA}$ always.}
    \label{fig:age_SN_vs_galaxy}
\end{figure}

The cosmological interpretation proposed by \Son relies on a key assumption, articulated by \Chung, that ``on average, the age differences between the stellar populations and SN~Ia progenitors in host galaxies are not significant.'' Under this assumption, the empirical HR--age slope measured with mass-weighted galaxy ages is treated as if it can be applied directly to progenitor ages. This step is only valid if it can be explicitly shown that $t_\mathrm{MWA}\approx t_\mathrm{prog}$ for the relevant host population.

Our data demonstrate that this approximation does not hold in general. The bimodality in Figure~\ref{fig:age_histogram} already indicates that progenitor ages separate strongly by host type, whereas mass-weighted galaxy ages do not uniquely encode the recent star formation that dominates the SN Ia production probability. Figure~\ref{fig:age_SN_vs_galaxy} makes this explicit by comparing $t_\mathrm{MWA}$ to the expected progenitor age $\mathbb{E}[t_\text{prog}]$. For star-forming hosts, $\mathbb{E}[t_\text{prog}]$ remains young ($\lesssim4\,\mathrm{Gyr}$) over a wide range of $t_\mathrm{MWA}$ (up to $\sim8\,\mathrm{Gyr}$), demonstrating offsets of up to $\sim4\,\mathrm{Gyr}$ between progenitor and galaxy ages.

For quiescent galaxies, progenitor ages span a wider range ($4\lesssim \mathbb{E}[t_\text{prog}] \lesssim 10\ \mathrm{Gyr}$) despite a comparatively narrow range of old galaxy ages ($7\lesssim t_\mathrm{MWA} \lesssim 10\ \mathrm{Gyr}$), and the mapping between $t_\mathrm{MWA}$ and $\mathbb{E}[t_\text{prog}]$ is strongly nonlinear, as for star-forming galaxies.

This behavior follows naturally from the SN~Ia DTD: because the DTD approximately scales as $t^{-1}$, even a modest episode of recent star formation in an otherwise old galaxy can dominate the delay-time-weighted explosion probability. As a result, the observed SN~Ia is likely to arise from a comparatively young progenitor population, regardless of the dominant stellar-mass component of the host (Fig.~\ref{fig:sfh_to_age_demo}). Figure~\ref{fig:age_SN_vs_galaxy} demonstrates that, when applying a standard DTD to empirically derived multiwavelength SFHs of SN~Ia hosts, the inferred delay-time-weighted progenitor age satisfies $\mathbb{E}[t_\text{prog}] < t_{\rm MWA}$ across the sample\footnote[1]{
A small fraction ($<3\%$) of the fitted SFHs exhibit a significant burst in star formation within the most recent $\sim100$~Myr, which results in $\mathbb{E}[t_\text{prog}]\lesssim500$~Myr, shown as a bimodal feature in low-mass SF group in Fig.~\ref{fig:age_SN_vs_galaxy}. Determining whether this is a physical indication for a separate progenitor channel or due to systematics in the SFH model requires further study, but our results are insensitive to these populations because (1) this is only applicable to $<3\%$ of the total hosts, and (2) a completely independent, binned SFH model (see Appendix~\ref{sec:app_iyer}) yields the progenitor age histogram consistent with our baseline result.}. 
Consequently, using $t_{\rm MWA}$ as a direct proxy for progenitor age, as assumed in \Son, systematically alters the inferred age-luminosity scaling. The implications of this distinction for redshift-dependent standardization corrections are discussed in Sec.~\ref{sec:cosmo}.

\section{On the Origin of the Discrepancy with S25} \label{sec:discussion}
In the previous section, we demonstrated that the empirically derived progenitor-age distributions --- and the resulting biases in Hubble residuals --- differ substantially from those inferred by \Chung\ and \Son. A central reason for this discrepancy is that progenitor age cannot be directly equated with the host galaxy’s mass-weighted stellar age, as discussed in Sec.~\ref{sec:age_vs_age}. Here, we examine additional differences in the assumptions and methodological choices underlying the two analyses. Specifically, we discuss
(i) the necessity of IR coverage in constraining galaxy and progenitor ages; 
(ii) selection effects in SN~Ia datasets, which likely render the redshift-dependent evolution of SN~Ia progenitor populations significantly more complex than assumed by \Son; and
(iii) uncertainties in the CSFH, which permit a broader range of predicted $z=0$ SPADs than considered by \Son. 
We argue that realistic bias estimates for cosmological applications of SN~Ia datasets require empirical and direct measurements of host-galaxy properties (e.g., mass, SFH) across redshift, rather than simplified modeling assumptions. Neglecting these effects can lead to mischaracterization of the inferred progenitor-age bias.

\subsection{UV--Optical Photometry is Insufficient for Constraining Age} \label{sec:need_mir}

\begin{figure}
    \centering
    \includegraphics[width=\columnwidth]{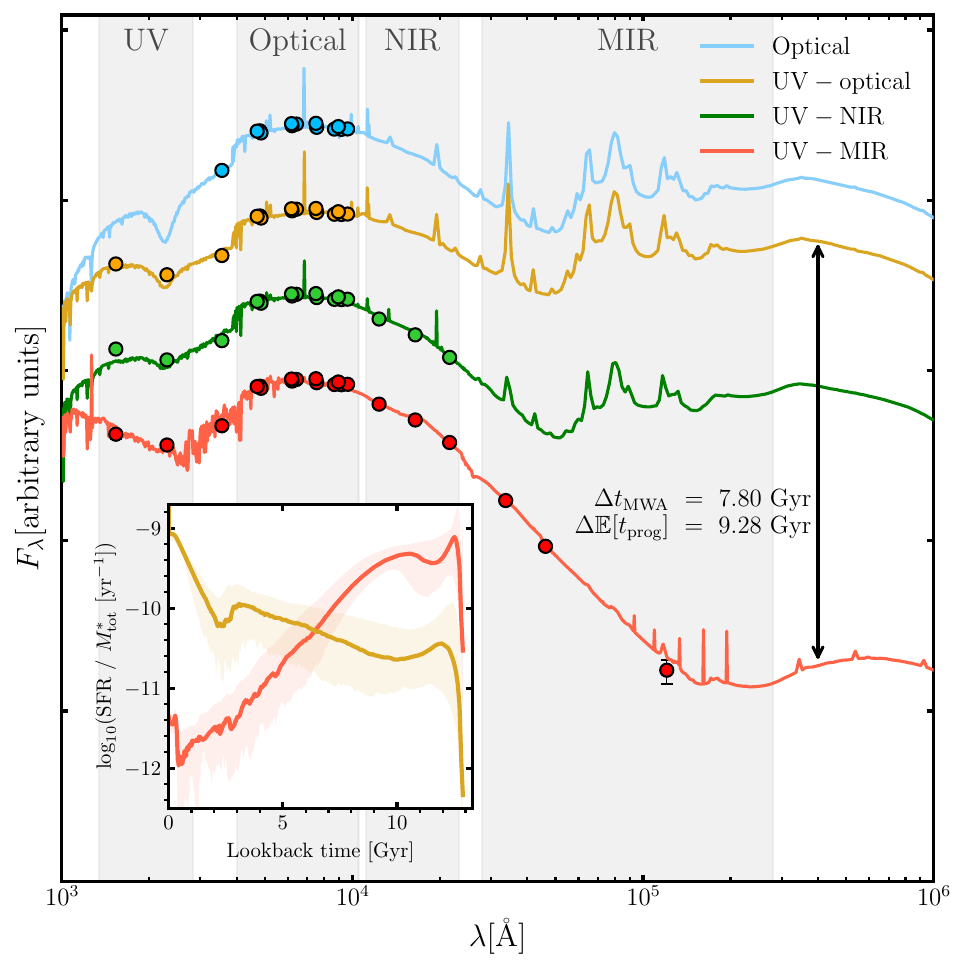}
    \caption{Best-fitting SED of the host galaxy of SN~2024vjb inferred using different subsets of the available photometric data. Colored curves show the posterior median SED obtained when fitting optical-only photometry (blue), UV--optical (yellow), UV--NIR (green), and UV--MIR (red), with colored points indicating the observed photometric measurements used in each case. Each instance is offset to improve clarity. Shaded regions denote the approximate wavelength ranges corresponding to the UV, optical, NIR, and MIR. The inset panel shows the corresponding inferred SFHs for the UV--optical and UV--MIR fits, illustrating how the inclusion of MIR data suppresses spurious recent star formation and favors an older stellar population. The annotated values (next to the vertical double arrow) indicate the change in mass-weighted stellar age, $\Delta t_\mathrm{MWA}=7.80\ \mathrm{Gyr}$, and the resulting change in the inferred SN Ia progenitor age, $\Delta \mathbb{E}[t_\mathrm{prog}]=9.28\ \mathrm{Gyr}$, when MIR photometry is included. This figure demonstrates how optical and UV--optical SED fitting alone can bias inferred galaxy and progenitor ages, and how MIR constraints are required to break the age-dust degeneracy.}
    \label{fig:change_phot_case_study}
\end{figure}

The SED of a galaxy encodes a complex interplay between stellar populations of different ages, dust attenuation, and dust reemission across a wide range of wavelengths \citep[e.g.,][]{Conroy2013}. Fits based solely on optical photometry are therefore intrinsically limited, as they suffer from strong degeneracies between stellar age, metallicity, and dust extinction \citep{Worthey1994, Papovich2001, Gallazzi2005}. The inclusion of UV data provides additional leverage on recent star formation \citep{Kennicutt1998, Salim2007}, but UV emission is itself highly susceptible to dust attenuation \citep[e.g.,][]{Calzetti2000}, such that galaxies with intense, dust-obscured star formation can exhibit UV fluxes comparable to those of systems with only modest, unobscured star formation. 

Consequently, SED fitting that excludes MIR or FIR data --- like that performed by \cite{Gupta2011}, \cite{Rose_2019_age}, and \Chung\ --- cannot reliably distinguish heavily obscured star-forming galaxies from genuinely quiescent hosts \citep{daCunha2008, Salim2016}, and the resulting age estimates can be strongly prior- and model-dependent. For this reason, we emphasize (Sec. \ref{sec:SFH}) the importance of MIR and FIR data, which directly trace dust-reprocessed starlight and provide essential, complementary constraints on the total star-formation activity. This additional information substantially reduces the dust-driven ambiguity in inferred recent SFHs, as illustrated in Figure \ref{fig:change_phot_case_study}. 

The case study of SN~2024vjb shown in Figure~\ref{fig:change_phot_case_study} presents four SED fits to the same galaxy, each using a different subset of photometric data. It illustrates one example whereby in the absence of MIR data, an intrinsically old, quiescent galaxy can be misidentified as dusty and star-forming. The recent spurious star formation inferred from the UV--optical fit shifts the mean $\mathbb{E}[t_\mathrm{prog}]$ to be $9.28\ \mathrm{Gyr}$ younger than inferred when the full UV--MIR wavelength range is considered; full posteriors for this example are presented in Appendix \ref{sec:app_bagpipes}. The converse misclassification --- where a genuinely young, star-forming galaxy is interpreted as old and quiescent, resulting in a spuriously older SN Ia progenitor --- is also plausible, as both scenarios reproduce extremely similar observed colors and are therefore essentially statistically indistinguishable without additional wavelength coverage. When comparing SED fits to UV--optical and UV--MIR data across the full TITAN sample, the progenitor-age bias induced by this degeneracy can be as large as $\sim 10\ \mathrm{Gyr}$. 

Against this backdrop, the host-galaxy ages reported by \Chung should be treated with caution, as they are derived from SED fits lacking the MIR (or FIR at higher redshift) data required to directly trace dust-reprocessed star formation. Without such constraints, inferred SFHs and stellar ages cannot be robustly interpreted as physical quantities --- nor, as demonstrated in Sec. \ref{sec:age_vs_age}, as reliable proxies for SN~Ia progenitor age --- and they should not be used to motivate age-dependent corrections to SN~Ia luminosities. These findings also underscore the need to rederive the HR--age relation using improved age estimates, which we defer to future work once TITAN Hubble residuals are finalized.

\subsection{Selection Effects} \label{sec:selection_effects}
The observed SPAD is not a direct tracer of the intrinsic product between the CSFH and the DTD, but is instead shaped by survey selection effects, follow-up strategy, and data-quality requirements. In this work, we limit ourselves to a qualitative assessment of the dominant biases, deferring a full forward-modeling of the TITAN selection function to future work (Tweddle et al., in prep.). 

The primary selection criteria for the TITAN bronze sample (Murakami et al., in prep.)  --- secure spectroscopic classification and host-galaxy identification --- introduce competing effects. The spectroscopic classification of SNe preferentially selects brighter, longer-lived SNe \citep[e.g.,][]{Perrett2010}, which are more common in star-forming environments \citep[e.g.,][]{DAndrea2011} and thus bias the sample toward younger progenitors. In contrast, requiring robust host association disfavors faint, low-mass galaxies \citep[e.g.,][]{Gupta2016}, which are more likely to host younger stellar populations (Fig.~\ref{fig:age_histogram}), thereby biasing the sample toward older progenitors.

\begin{figure}
    \centering
    \includegraphics[width=\linewidth]{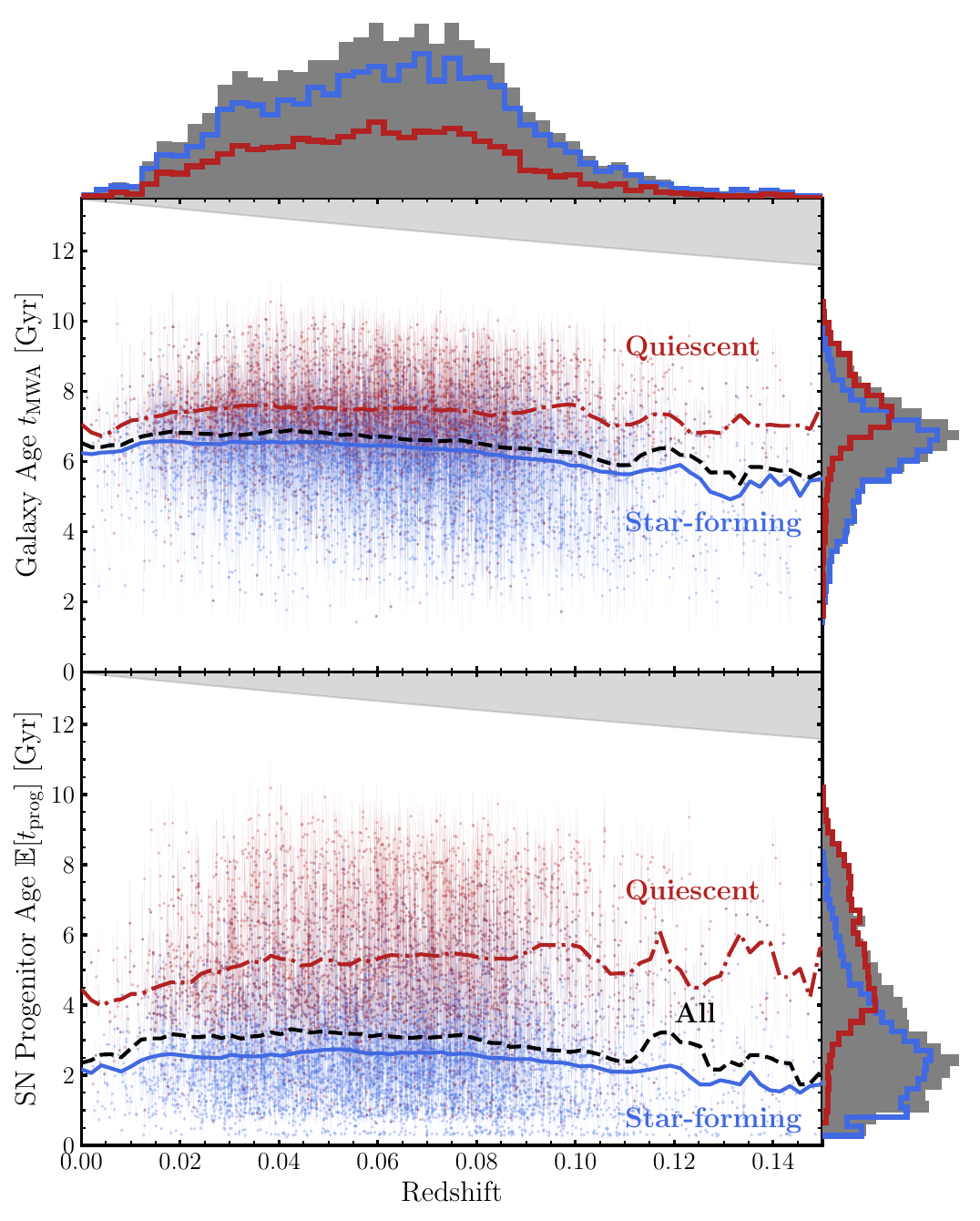}
    \caption{Galaxy-mass-weighted age and delay-time-weighted progenitor age as a function of redshift in the TITAN sample.
    \textit{Top:} Host-galaxy mass-weighted stellar age ($t_\mathrm{MWA}$) versus redshift for quiescent (red) and star-forming (blue) systems; all galaxies are shown as a black line. Solid and dashed curves show the running median in redshift bins. 
    The filled gray region indicates prohibited progenitor ages exceeding the age of the Universe at a given redshift. Histograms on the right show the marginal age distributions for all TITAN hosts, quiescent hosts, and star-forming hosts. \textit{Bottom:} Corresponding redshift evolution of expectation values for SN~Ia progenitor ages ($\mathbb{E}[t_\mathrm{prog}]$).}
    \label{fig:age_vs_redshift}
\end{figure}

Additional selection cuts imposed by cosmological analyses (e.g., the TITAN gold sample, further requiring a cosmology-grade light-curve and host-galaxy spectroscopic redshift; Murakami et al., in prep.) further complicate this picture. Requirements on light-curve quality favor intrinsically brighter, high-$x_1$ SNe, again skewing the sample toward younger progenitors. At the same time, the need for spectroscopic host-galaxy redshifts introduces a bias toward more massive, evolved systems, and hence older progenitors. However, at fixed galaxy mass, spectroscopic redshifts are more readily obtained for star-forming galaxies owing to their strong, narrow emission lines, partially counteracting this trend by favoring younger systems. These effects act in opposing directions and are strongly redshift dependent, as only the brightest SNe and most massive hosts are retained at higher redshift. As a result, the observed SPAD can exhibit an apparent redshift evolution that is driven by selection rather than intrinsic changes in the progenitor population. The modeling of DES selection effects by \Wiseman suggests that the net evolution in progenitor age is minimal over $0.1 \lesssim z \lesssim 1.2$; however, this conclusion is survey-dependent and may vary with differing selection functions. We observe no statistically significant evolution in either galaxy mass-weighted age or expected SN Ia progenitor age in the narrow redshift range of $z\approx0$ to $z\approx0.15$, as shown in Figure~\ref{fig:age_vs_redshift}.

The interplay of competing selection effects is evident within TITAN data: restricting to the cosmological / gold TITAN sample , increases the median progenitor age by $\sim 200\ \mathrm{Myr}$ (from $1.9$ to $2.1\ \mathrm{Gyr}$), indicating that the net effect of additional selection cuts is to enhance the contribution of older systems.  The magnitude of this shift is comparable to the evolution predicted by several progenitor models across the redshift range probed here, implying that unmodeled selection effects could be misinterpreted as physical evolution. A robust interpretation therefore requires joint modeling of the survey selection function, host-galaxy demographics, and the mapping between progenitor age and SN observables, which we leave to future work.

\subsection{Sensitivity to Uncertainties in the CSFH}

\begin{figure}
    \centering
    \includegraphics[width=\columnwidth]{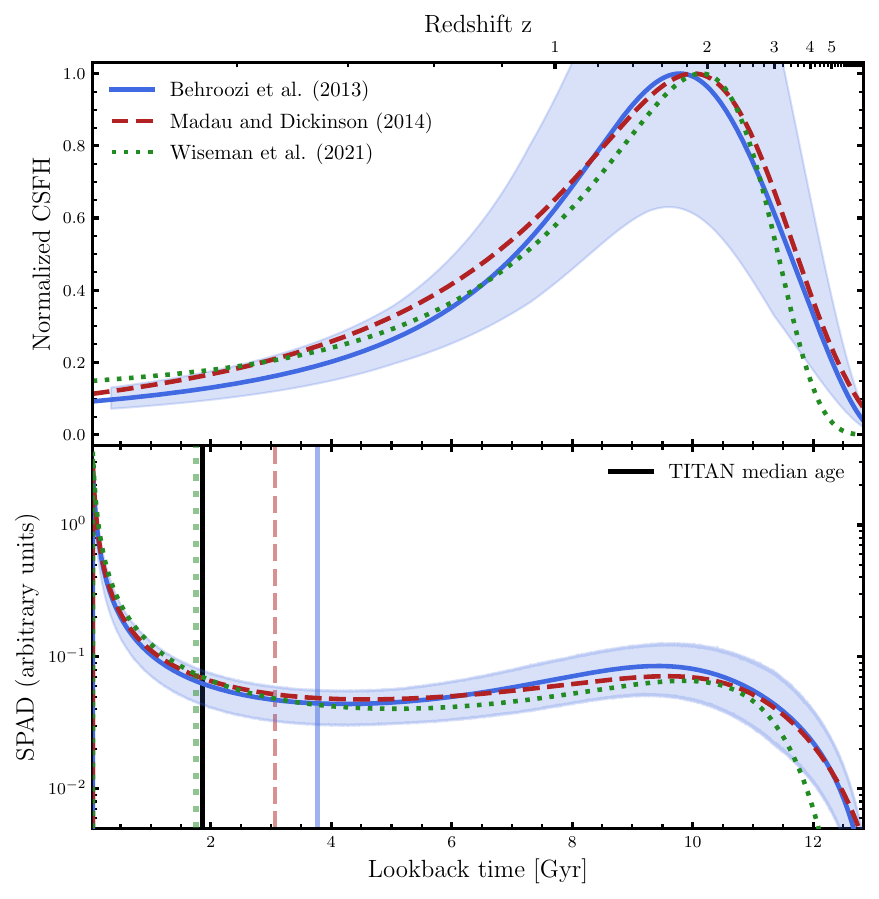}
    \caption{Impact of the assumed cosmic star-formation history (CSFH) on the predicted SN~Ia progenitor-age distribution (SPAD). \textit{Top:} Comparison of normalized CSFH models from \Behroozi and \Madau, alongside the reconstruction from \cite{Wiseman2021}, as a function of redshift (top axis) and lookback time (bottom axis). The shaded region shows the systematic uncertainty on the \Behroozi CSFH, using the estimates from Table 7 of \Behroozi. \textit{Bottom:} Corresponding $z=0$ SPADs obtained by multiplying each CSFH with a DTD having $t_{\rm min}=40$~Myr and $\beta_\mathrm{DTD}=-1.07$. Vertical lines indicate the median progenitor age for each model, while the solid black line marks the median of the TITAN data. Despite relatively small differences in the CSFHs and overall shape of the SPADs, the predicted medians vary significantly, illustrating the sensitivity of the model predictions to the assumed CSFH.}
    \label{fig:diff_csfh}
\end{figure}

Under the assumption of a \citet[hereafter \Behroozi]{Behroozi2013} CSFH (as adopted by \citealt{Childress_2014_age} and \Son) and DTD parameters $t_\mathrm{min}=40\ \mathrm{Myr},\ \beta_\mathrm{DTD}=-1.07$, the predicted median of the SPAD at $z=0$ is $3.8\ \mathrm{Gyr}$, compared to our measured value of $1.9\ \mathrm{Gyr}$. However, this prediction is highly sensitive to the assumed CSFH. Even modest differences in the shape of the CSFH can produce large shifts in the inferred SPAD, owing to the steep $t^{-1}$-like dependence of the DTD.

n Figure~\ref{fig:diff_csfh}, we compare predictions obtained using the measured cosmic star formation history of \citet[hereafter \Madau]{Madau2014} and the reconstructed CSFH from \cite{Wiseman2021}. While the differences between these CSFH models are relatively subtle in appearance --- at the level of $\sim 20\%$ in the recent ($\lesssim 5$ Gyr) SFR --- the resulting medians of the $z=0$ SPADs differ significantly. In particular, the median progenitor age shifts from $3.8\ \mathrm{Gyr}$ for \Behroozi, to $3.1\ \mathrm{Gyr}$ for \Madau, and down to $1.8\ \mathrm{Gyr}$ for \cite{Wiseman2021}, in good agreement with our measurement. Therefore, the apparent tension between our measured SPAD and predictions based on the \Behroozi CSFH does not necessarily imply a discrepancy in the choice of DTD or our SED modeling, but may instead reflect uncertainties in the low-redshift CSFH.

Furthermore, the TITAN SPAD is constructed from SNe spanning the redshift range $0 \lesssim z \lesssim 0.15$, rather than representing a strictly $z=0$ measurement. As such, it implicitly averages over a finite redshift range, incorporating both any underlying evolution in the progenitor-age distribution and the impact of redshift-dependent selection effects. Because progenitor ages are expected to decrease with increasing redshift, this averaging will bias the inferred SPAD toward younger ages relative to a true $z=0$ prediction. However as discussed above (Fig.~\ref{fig:age_vs_redshift}), we observe no statistically significant evolution in either galaxy-mass-weighted age or expected progenitor age across this range, indicating that evolutionary and selection effects at least partially offset one another. While likely subdominant to uncertainties in the CSFH, this further complicates direct comparisons between observations and simple $z=0$ models.

Taken together, these results demonstrate that the predicted SPAD is not uniquely determined by the choice of DTD, but instead reflects a coupled dependence on both the DTD and the assumed CSFH. This degeneracy implies that discrepancies between observed and predicted progenitor-age distributions cannot be straightforwardly interpreted without simultaneously constraining the underlying SFH. In this context, the SPAD itself may provide a complementary probe of the recent CSFH, albeit one that is intrinsically convolved with the DTD and subject to observational selection effects.

\section{Implications for Cosmology} \label{sec:cosmo}

In this section, we combine our measured SN~Ia progenitor-age distribution and derived $t_\mathrm{MWA}$–$t_\mathrm{prog}$ relation with literature $\mathrm{HR}$–$t_\mathrm{MWA}$ constraints to assess the expected impact on cosmology. 
We first estimate the redshift-dependent bias arising from neglecting age-related corrections by modeling the evolution of the mean progenitor age. We then revisit the expected magnitude of age-dependent effects within the \Son\ framework, with cautionary remarks and discussions on a more realistic pathway toward quantifying the age-dependent bias in cosmological analysis.

\subsection{Expected Progenitor-Age-Dependent Biases}
\label{sec:age-HR-bias}
Recent cosmological measurements with state-of-the-art datasets, such as Pantheon+ \citep{Brout2019} or DES \citep{DES_2024_SNIa_Y5}, include a mass-step correction that accounts for the majority of the observed host-dependent trends, including correlations between HRs and galaxy-mass-weighted age, $t_\mathrm{MWA}$. In \Wiseman, the HR--$t_\mathrm{MWA}$ slope is reevaluated after applying the standard mass-step and bias corrections, and is found to be  $-0.007^{+0.012}_{-0.014}\ \mathrm{mag\ Gyr^{-1}}$, consistent with no significant residual age dependence. This is in contrast to the $-0.022^{+0.011}_{-0.009}\ \mathrm{mag\ Gyr^{-1}}$ slope measured prior to host or bias corrections, which is derived by \Wiseman from a reanalysis of the \Chung/\Son sample before applying host or bias corrections.

\begin{figure}
    \centering
    \includegraphics[width=\linewidth]{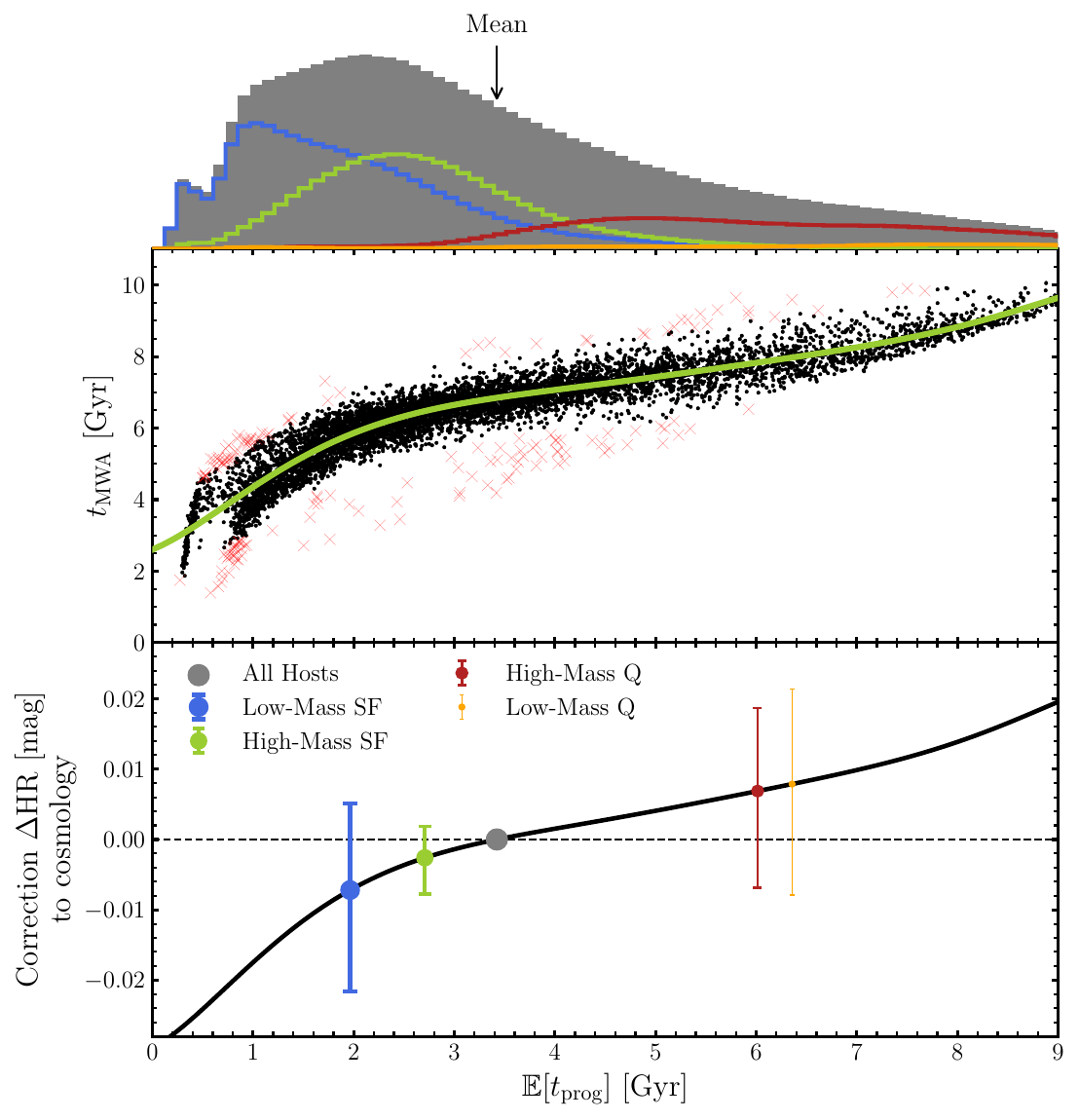}
    \caption{Inferred age-dependent bias from the \Wiseman HR--$t_\mathrm{MWA}$ slope and the $t_\mathrm{MWA}$--$\mathbb{E}[t_\mathrm{prog}]$ relation derived in this work. In the limiting case where all SN~Ia hosts converge to low-mass, star-forming galaxies at high redshift, the model predicts a shift of $-0.007^{+0.012}_{-0.014}$~mag in the mean HR between low- and high-redshift samples. \textit{Top:} The progenitor-age--galaxy-age relation. We fit a polynomial with iterative outlier rejection to obtain the representative transformation. \textit{Bottom:} The predicted HR--$\mathbb{E}[t_\mathrm{prog}]$ trend, obtained by applying the $t_\mathrm{MWA}$--$\mathbb{E}[t_\mathrm{prog}]$ relation in the top panel to the HR--$t_\mathrm{MWA}$ slope measured in \Wiseman. Each point represents the mean of the measured progenitor ages for different host types and their corresponding $\Delta\mathrm{HR}$ values, with error bars reflecting the uncertainty on the \Wiseman HR--$t_\mathrm{MWA}$ slope. The marker size represents the relative contribution of each population to the TITAN dataset.}
    \label{fig:tSN_vs_tGal_vs_HR}
\end{figure}

Here, we use the mass-step- and bias-corrected HR--$t_{\rm MWA}$ slope from \Wiseman to estimate the magnitude of any progenitor-age-dependent bias by mapping host-galaxy mass-weighted age to progenitor age using the empirically derived $t_{\rm MWA}$--$\mathbb{E}[t_{\rm prog}]$ relation shown in Figure~\ref{fig:age_SN_vs_galaxy}. This approach yields a physically motivated HR--$\mathbb{E}[t_{\rm prog}]$ relation and is more realistic than assuming that an observed HR--$t_{\rm MWA}$ correlation can be directly applied to progenitor age (and its redshift evolution), as discussed in Sec.~\ref{sec:age_vs_age}.

The resulting HR--$\mathbb{E}[t_{\rm prog}]$ relation is shown in Figure~\ref{fig:tSN_vs_tGal_vs_HR}. The top panel illustrates the polynomial transformation between $t_{\rm MWA}$ and $\mathbb{E}[t_{\rm prog}]$ inferred from the TITAN host sample, while the bottom panel shows the corresponding impact on the Hubble residuals. 
A key distinction between this empirically derived HR--$\mathbb{E}[t_{\rm prog}]$ relation and the \Son\ model --- where $\Delta\mathrm{HR}\propto t_{\rm MWA}\approx \mathbb{E}[t_{\rm prog}]$ --- is not only the order-of-magnitude reduction in the overall scale of $\Delta\mathrm{HR}$, but also that for older stellar populations ($\mathbb{E}[t_\mathrm{prog}]\gtrsim4\ \mathrm{Gyr}$), additional increases in host-galaxy age produce progressively smaller changes in progenitor age and therefore in the inferred Hubble residual. As a result, age differences among older populations have a diminishing impact on $\Delta\mathrm{HR}$, in contrast to the linear scaling assumed by \Son.

At higher redshift, typical galaxy masses decrease and the stellar populations of quiescent galaxies systematically become younger, leading to correspondingly younger SN~Ia progenitors. In the limit of sufficiently high redshift, the progenitor-age distribution in quiescent systems is expected to converge toward that of low-mass, star-forming galaxies. Because the stellar populations of star-forming galaxies are continually rejuvenated, their characteristic progenitor age distribution is not expected to evolve strongly with redshift and therefore represents the youngest baseline distribution of SN~Ia progenitor ages \citep{Childress_2014_age}.

We therefore expect that, at all epochs when the age of the Universe exceeds this timescale, the majority of SNe~Ia arise from systems with characteristic progenitor ages of $\sim 2.0\ \mathrm{Gyr}$. Our data therefore predict an evolution in the expected mean SN Ia progenitor age of $\sim 1.5\ \mathrm{Gyr}$ over cosmic time (or an evolution in mean $t_\mathrm{MWA}$ of $\sim1\ \mathrm{Gyr}$), corresponding to a shift from the present-day mean of 3.5~Gyr to the low-mass, star-forming limit (shown by the gray and blue points, respectively, in Figure~\ref{fig:tSN_vs_tGal_vs_HR}).  Applying the empirically derived, host- and bias-corrected HR--$\mathbb{E}[t_{\rm prog}]$ relation under these assumptions yields a maximum redshift-dependent, progenitor-age-driven Hubble residual bias of $-0.007^{+0.012}_{-0.014}$~mag. We thus find that any redshift-dependent bias induced by progenitor-age evolution is consistent with zero.

\begin{figure*}
    \centering
    \includegraphics[width=\linewidth]{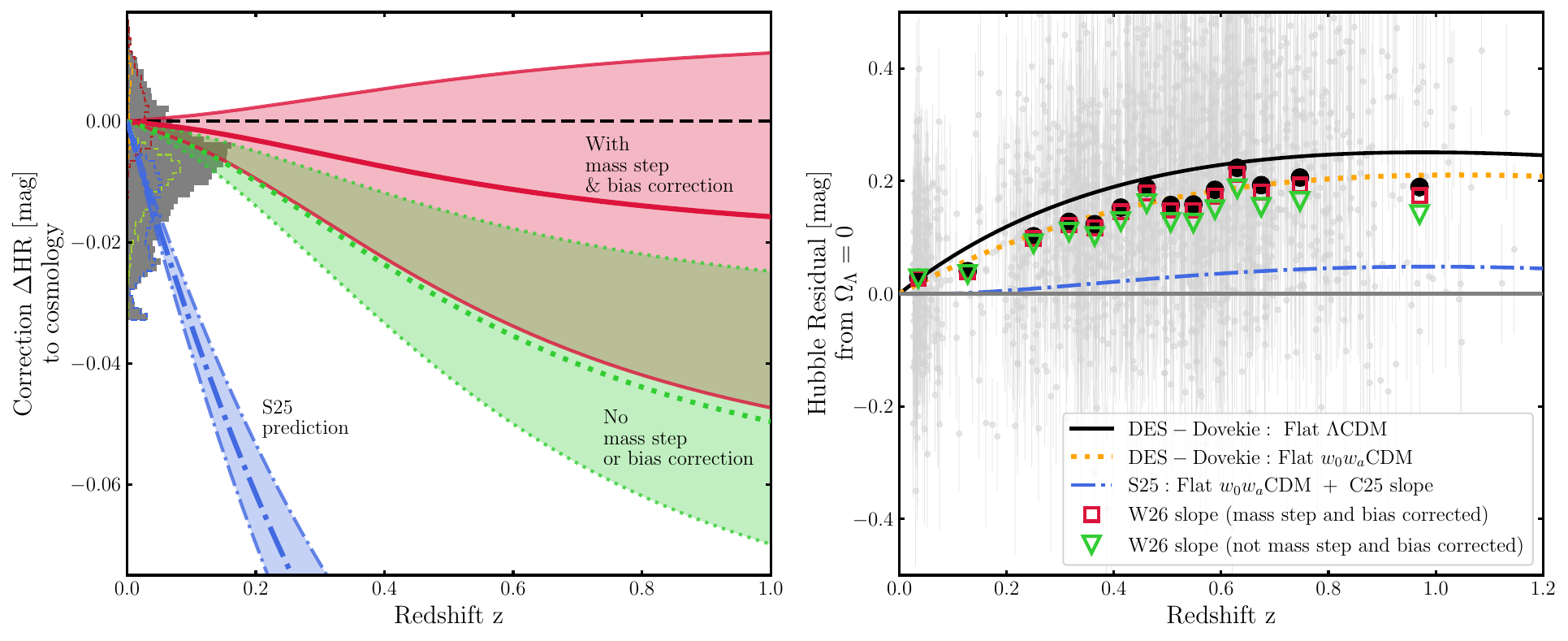}
    \caption{Predicted redshift-dependent bias from redshift evolution of the median of the SPAD and its impact on cosmological residuals of SNe Ia.
    \textit{Left:} Expected redshift evolution of the Hubble residual correction, $\Delta$HR, induced by progenitor-age evolution. The signal is computed by combining the HR--$t_{\rm MWA}$ slope from \Wiseman\ with the empirically derived $t_{\rm MWA}$--$\mathbb{E}[t_\mathrm{prog}]$ mapping from the TITAN host sample, together with the predicted progenitor-age evolution implied by the \cite{Wiseman2021} CSFH reconstruction. The solid (red) curve shows the prediction including standard host-mass and bias corrections, while the dotted (green) curve illustrates the case without these corrections. The dash-dotted (blue) curve displays the evolution predicted by the \Son\ model, which assumes a linear mapping between galaxy age and progenitor age and a steeper HR--age relation. Shaded bands show the $1\sigma$ propagated uncertainty from the HR--age slope. The dashed black line denotes no correction. The vertical histogram at $z=0$ illustrates the distribution of relative HR corrections for the TITAN sample under the scaling model described in the text. \textit{Right:} Impact of these corrections on binned SN~Ia Hubble residuals relative to an Einstein--de Sitter cosmology. Gray points show individual DES5YR SNe, while black points indicate binned measurements (equal-number bins) from the DES-Dovekie reanalysis \citep{Popovic2025}, shifted by a constant offset to align with the best-fit DES-Dovekie $w_0w_a\mathrm{CDM}$ cosmology. The solid black and orange dotted curves show the best-fit DES-Dovekie flat $\Lambda$CDM and flat $w_0w_a\mathrm{CDM}$ cosmologies, respectively, while the blue dash-dotted curve represents the \Son\ prediction. Colored markers indicate the effect of applying the progenitor-age evolution model derived here, with (red squares) and without (green inverted triangles) host-mass and bias corrections. All cosmological models are fit to SN+BAO+CMB data.
    }

    \label{fig:HR_vs_z}
\end{figure*}

\subsection{Updated Cosmological Implications from S25 Methodology}
\label{sec:S25-reanalysis}
In contrast to the methodology presented in the previous section, where the evolution in the expected SN~Ia progenitor age is used to quantify the magnitude of any progenitor-age-driven bias, \Son instead consider the redshift evolution of the median of the SPAD. This distinction is nontrivial: the median does not, in general, trace the expectation value that enters the HR--age relation, and therefore does not directly map onto the predicted bias. This choice of statistic is particularly important given that cosmological analyses typically adopt Gaussian likelihoods \cite[e.g.,][]{DES_2024_SNIa_Y5}, for which parameter inference is primarily sensitive to the mean of the underlying population. For strongly skewed distributions such as the SPAD, the mean and median can differ substantially, reflecting different aspects of the population. Consequently, using the median as a proxy for the relevant progenitor age can lead to a mischaracterization of the expected bias.
Nevertheless, in this section, we show that, even within this alternative framework, the inferred redshift-dependent progenitor-age bias remains consistent with zero.

To adopt this methodology, we require a model for the redshift evolution of the SPAD median, as our data do not span a sufficiently wide redshift range to measure this directly. We therefore adopt the evolution predicted by the \cite{Wiseman2021} CSFH model, whose $z=0$ SPAD median is consistent with that measured from the TITAN sample (Fig.~\ref{fig:diff_csfh}). We note that this approach neglects selection effects and instead reflects the evolution of the intrinsic SPAD; however, this is sufficient for the purposes of the arguments that follow.

In the left panel of Figure~\ref{fig:HR_vs_z}, we apply the host- and bias-corrected HR--$\mathbb{E}[t_\mathrm{prog}]$ relation to the predicted evolution of the SPAD median. This yields a net redshift-dependent bias in the Hubble residual of $-0.016_{-0.029}^{+0.027}$~mag by $z=1$. Neglecting host-mass and bias corrections increases the inferred bias, as expected, to $-0.050^{+0.025}_{-0.020}$~mag. In both cases, the resulting impact on cosmology is substantially smaller than the $-0.159_{-0.021}^{+0.022}$~mag reported by \Son, corresponding to $5.3\ \mathrm{Gyr}$ of evolution in the SPAD median.

The right panel of Figure~\ref{fig:HR_vs_z} shows the corresponding impact of the redshift-dependent bias (derived in the left panel) on the Hubble diagram. Specifically, we propagate the redshift-dependent $\Delta\mathrm{HR}(z)$ inferred from our SPAD evolution model into the DES-Dovekie SN~Ia sample \citep{Popovic2025}, shown relative to an Einstein--de Sitter cosmology. The DES-Dovekie binned data (black points) and the best-fit flat $\Lambda$CDM (black curve) and $w_0w_a\mathrm{CDM}$ (orange dotted curve) cosmologies to the DES5YR+CMB+BAO data are displayed for reference, as well as the best-fit model from \Son (blue dash-dotted curve), which incorporates the \Chung HR--$t_\mathrm{MWA}$ slope.

When the host- and bias-corrected HR--$t_{\rm prog}$ relation derived from \Wiseman\ is applied, the induced redshift-dependent offset is at the level of a few mmag across the full redshift range. This shift is small compared to both the intrinsic scatter of individual SNe~Ia ($\sim0.15$~mag) and the statistical uncertainties of current high-redshift binned measurements. If host and bias corrections are neglected, the induced offset increases but remains substantially smaller than the (up to) $0.16$~mag deviations implied by the \Son\ model. The net effect on cosmological inference is therefore minimal, and we again find that any redshift-dependent bias induced by progenitor-age evolution is too small to materially alter current SN~Ia cosmological constraints.

\subsection{Future Approaches: From Estimates to Measurements}

It is important to note that the above analyses presented in Secs.~\ref{sec:age-HR-bias} and \ref{sec:S25-reanalysis} are order-of-magnitude estimates of biases in SN~Ia luminosity possible under the hypothesis proposed by \Son. 
As discussed in Sec.~\ref{sec:selection_effects}, the actual magnitude of age evolution over redshift is significantly dependent on selection effects, and one cannot determine the exact size of the bias and/or correct it without accounting for all possible effects of selection or evolution simultaneously. The model-dependent, idealistic and analytical approach as presented by Sec.~\ref{sec:age-HR-bias}, Sec.~\ref{sec:S25-reanalysis}, or \Son are useful for \textit{estimating} the size of the possible bias, but it alone does not enable direct updates to cosmological inferences.

In modern, robust cosmological analyses of SNe~Ia \citep[e.g.,][]{Brout_2022_PantheonPlus,DES_2024_SNIa_Y5}, bias corrections are often based on forward modeling and simulations; these simulations are based on actual survey design, observing conditions, telescope properties, application of quality cuts, as well as the estimated intrinsic distribution of SNe and host properties designed to match observations \citep[see, e.g.,][for an overview]{Kessler_Scolnic_2017_BBC}. 
A viable pathway to quantify the actual size of bias due to progenitor ages, therefore, is to implement a model to vary the SN~Ia progenitor age distribution over redshift, connect it to other observables (e.g., stretch), and simulate the selection effects in packages like \texttt{SNANA} \citep{Kessler2009}. Such simulations can then be compared to the actual observation of progenitor ages at a wide range of redshifts (not only at low $z$), and the age-HR slope can be determined simultaneously with other standardizations, such as the stretch-luminosity slope, color-luminosity slope, and mass-step. 

Achieving this requires that (i) we have a physically motivated, reliable understanding of the host dependency of SN~Ia progenitor age; (ii) we are able to obtain reliable age estimates for SN~Ia hosts at high redshift (see Sec.~\ref{sec:selection_effects} for discussion); and (iii) we can account for additional sources of systematics, especially covariance, that are introduced by the choice of DTD or CSFH. In particular, uncertainties in the observationally determined CSFH (see Fig.~\ref{fig:diff_csfh}) will likely yield a significant systematic covariance between SNe at different redshifts.

The closest existing analysis of this kind is \cite{Wiseman2022}, who forward-modeled the DES5YR sample with a physically motivated host–progenitor-age framework addressing (i) and (ii), and found that an additional age-dependent luminosity correction is disfavored once the standard mass-based bias corrections are applied. 
Until simulations that also address (iii), or Bayesian models that absorb the same effects (e.g., \citealt{Rubin2025, Boyd2026}), become available, any attempt to directly apply an age-based bias correction without addressing (i)–(iii) risks yielding misleading results. Nonetheless, independent lines of evidence, including the estimates in previous sections and the forward-modeled analysis of \cite{Wiseman2022}, all point to a small, statistically insignificant bias. 

\subsection{HR--Age Slope and the Hubble Constant}
\Son mention the possible implication of a progenitor-age effect on the Hubble constant (H$_0$) measurement, arguing that a possible mismatch of progenitor ages between the second rung (Cepheid-calibrated SNe~Ia) and the third rung (Hubble flow; HF) could result in a biased measurement of H$_0$. \cite{Riess_2022_SH0ESmain} address this potential issue by selectively using SNe from spiral hosts for their HF SN subsample. The differences in SN~Ia progenitor ages between rungs discussed by \Son would require a significant contamination of quiescent hosts within morphologically classified spiral groups.

\cite{Paspaliaris_2023} study star formation in local galaxies of various morphological types using NUV--FIR photometry and spectroscopic classifications. They find that, for Sa-Scd type (i.e., spiral galaxies), 8\% are classified as quiescent. 
Assuming that 92\% of spiral hosts for Hubble-flow SNe are star-forming, the impact of contamination on the mean progenitor age is minimal: random draws of progenitor ages from our distribution in Figure~\ref{fig:age_histogram} yields a $0.25$~Gyr shift in the mean age when 8\% quiescent contamination is added.
Using the $0.03$~mag Gyr$^{-1}$ HR--age slope from \Son, this results in a $<0.01$~mag shift in HRs, and with a more realistic re-evaluated slope by \Wiseman of $0.007$~mag/Gyr, the impact on the SN~Ia residual is 0.00175 mag. This result is an order of magnitude smaller than the 2--3~Gyr shift suggested by \Son.

Realistically, the Hubble flow SNe could be younger on average. In the SH0ES \citep{Riess_2022_SH0ESmain} sample, 65\% of second-rung SNe belong to high-mass ($\log_{10}(M_*/M_\odot)>10$) galaxies, while the number drops to 48\% for SNe in the third rung. As discussed in Sec.~\ref{sec:age}, high-mass hosts produce SNe slightly older than low-mass hosts by $\lesssim1$~Gyr within the star-forming group. 
Similarly to the calculation above, drawing from high- and low-mass hosts with the 65-35 and 48-52 ratios for the second and third rung, respectively, our TITAN age distribution predicts the progenitors of SNe in the third rung to be $\sim0.15$~Gyr younger on average than the calibrator hosts\footnote[2]{\cite{Riess_2022_SH0ESmain} measured the sSFR of SNe~Ia hosts, which provides a picture consistent with our discussion.}. This result shows a strong consistency in progenitor age between rungs, and its small, sub-Gyr level possible effect mostly cancels out with the quiescent contamination discussed above.

Combined with the \Wiseman HR--age slope, neither of these effects is significant at the $\sim0.01$~mag level.

\section{Conclusions} \label{sec:conclusion}

In this work, we presented a data-driven reassessment of whether SN~Ia progenitor-age evolution can generate a large, unmodeled, redshift-dependent bias in cosmology. Using the homogeneous TITAN low-redshift SN~Ia host-galaxy sample, comprising 6,983 hosts with broad UV--MIR photometric coverage, we constrained their SFHs using Gaussian-process reconstruction. Combining these SFHs with a literature SN~Ia DTD, we estimated delay-time-weighted progenitor ages for all events. We find that the local SN~Ia progenitor-age distribution is dominated by young systems, with a mean age of 3.5~Gyr. Our measurements imply only modest redshift evolution in the mean progenitor age of $\sim1.5$~Gyr, and a corresponding maximum age-driven Hubble-residual bias of $-0.007^{+0.012}_{-0.014}$~mag. We therefore find no statistically significant evidence for an additional redshift-dependent bias in cosmological analyses at current precision.

In summary, we find the following.
\begin{itemize}
    \item The inferred SPAD in the local Universe is strongly weighted toward young SN progenitors, with a primary peak at 2.2~Gyr (dominated by star-forming hosts) and a secondary component at 6.0~Gyr (arising predominantly from high-mass, quiescent hosts). The mean progenitor age in TITAN is 3.5~Gyr.
    \item The characteristic separation in progenitor age between (high-mass) star-forming and quiescent hosts is $3.30\pm0.03\,\mathrm{Gyr}$. An age--luminosity relation as strong as that claimed by \Son would therefore predict a post-standardization offset of $0.099 \pm 0.013$~mag between these populations --- substantially larger than the observed $0.037 \pm 0.019$~mag \citep{Wiseman2026}. This discrepancy directly disfavors such a strong intrinsic age dependence at the population level.
    \item Galaxy-mass-weighted stellar ages $t_\mathrm{MWA}$ are not interchangeable with SN Ia progenitor ages: $\mathbb{E}[t_\mathrm{prog}]$ remains young across a wide range of $t_\mathrm{MWA}$ for star-forming galaxies, and the mapping between $t_\mathrm{MWA}$ and $\mathbb{E}[t_\mathrm{prog}]$ is nonlinear, especially for quiescent systems. We find $\mathbb{E}[t_\mathrm{prog}] < t_\mathrm{MWA}$ always.
    \item Robust progenitor-age inference in the local universe requires constraining host-galaxy SFHs with MIR photometry to break the age-dust degeneracy; optical or UV--optical SED fitting can bias inferred SFHs and, in turn, progenitor ages by up to several Gyr. Correlations between HR and age should be subsequently reevaluated in light of improved age estimates.
    \item Selection effects are a critical component of any age-based cosmology test: the intrinsic CSFH$\times$DTD progenitor-age distribution is not directly observed, but is reshaped by survey discovery thresholds, light-curve quality cuts, and host-redshift requirements in redshift-dependent ways.
    \item Uncertainties in the CSFH propagate directly into predicted progenitor-age evolution; CSFH-informed estimates are therefore useful for scale-setting but are insufficient as standalone precision bias corrections.
    \item Using the host- and bias-corrected HR--$t_\mathrm{MWA}$ slope from \Wiseman together with our empirically derived $t_\mathrm{MWA}$--$\mathbb{E}[t_\mathrm{prog}]$ relation, we find that any plausible redshift-dependent bias from progenitor-age evolution is consistent with zero. This conclusion is unchanged whether one tracks evolution in the mean of $\mathbb{E}[t_\mathrm{prog}]$ or in the SPAD median (following the \Son methodology).

\end{itemize}

Overall, our results support a physically consistent picture in which SN~Ia progenitor ages are generally much younger than host-galaxy mass-weighted ages, and in which the observed host-dependent luminosity trends are already captured, to first order, by existing host-mass corrections. Looking ahead, the most informative next step is full forward modeling that jointly evolves progenitor-age distributions, light-curve populations, and survey selection across redshift, anchored by direct age constraints from low- and high-$z$ host samples. With this framework, future analyses can test progenitor-age-dependent standardization in a self-consistent way and place tighter empirical bounds on any residual age-driven cosmological bias for next-generation dark-energy measurements.

\begin{acknowledgements}
J.W.T. gratefully acknowledges support by the STFC through grant ST/Y509474/1. P.W. is grateful for the support STFC Ernest Rutherford Fellowship, grant ST/Z510269/1. D.O.J. was supported by NSF grants AST-2407632, AST-2429450, and AST-2510993, NASA grant 80NSSC24M0023, and {\it HST/JWST} grants HST-GO-17128.028 and JWST-GO-05324.031, awarded by the Space Telescope Science Institute (STScI), which is operated by the Association of Universities for Research in Astronomy, Inc., for NASA, under contract NAS5-26555.
SJS acknowledges funding from STFC Grants ST/Y001605/1, ST/X001253/1, a Royal Society Research Professorship and the Hintze Family Charitable Foundation. 
L.G. received financial support from CSIC, MCIN, and AEI 10.13039/501100011033 under projects PID2023-151307NB-I00, PIE 20215AT016, and CEX2020-001058-M. S.W.J.  acknowledges support from a Guggenheim Fellowship, DOE award DE-SC0010008, and NSF grant AST-2407567.  C.L. was funded through \textit{JWST} program grants JWST-GO-06541, JWST-GO-06585, and JWST-GO-05324. Y.S.M. and J.W.T. thank R. Wang and N. Redfewds for their support and helpful discussions.
A.V.F. is grateful for financial support from numerous donors.

\end{acknowledgements}

\appendix

\section{Sensitivity to the choice of SFH} \label{sec:app_iyer}

\begin{figure*}
    \centering
    \includegraphics[width=1\linewidth]{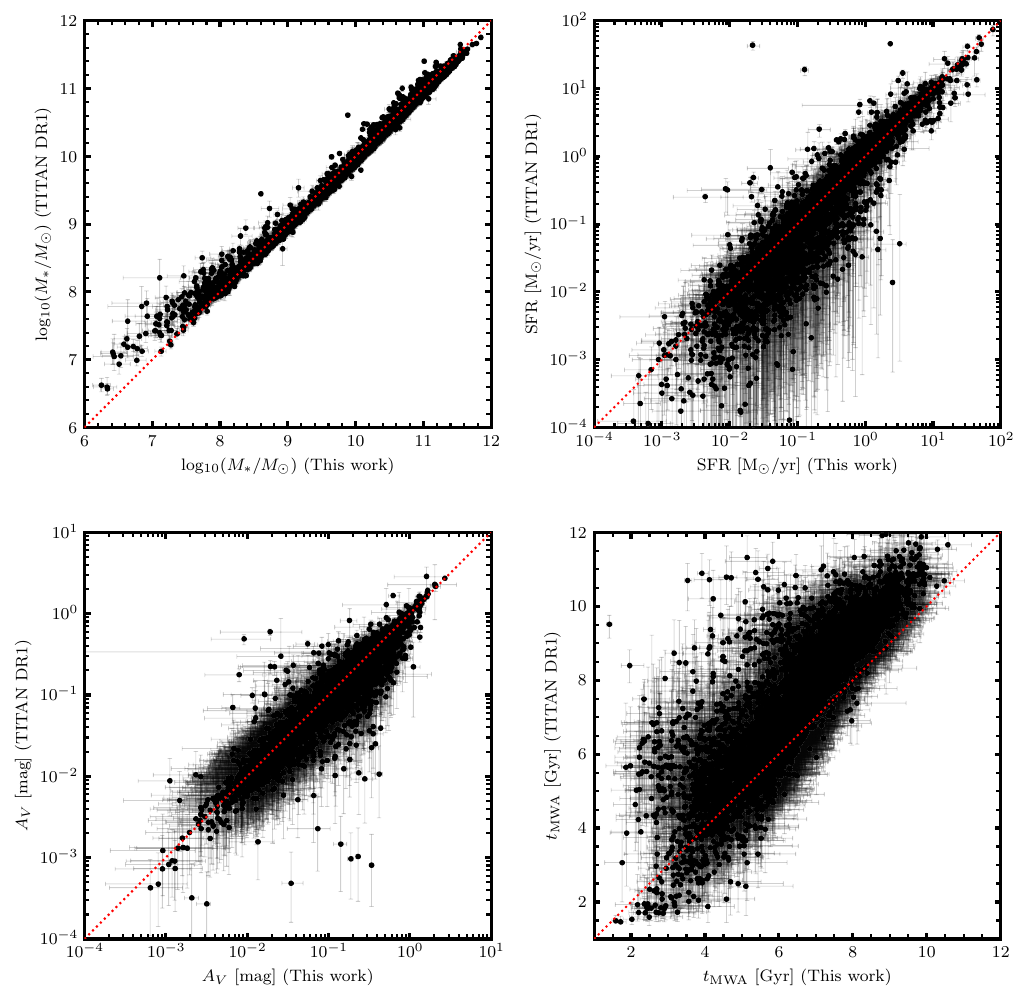}
    \caption{Comparison between fitted \texttt{Bagpipes} galaxy properties derived assuming an \cite{Iyer_2019_nonparamSFH} SFH (this work) to those derived assuming a \cite{Leja2019} nonparametric SFH with a continuity prior (as used in the TITAN DR1 data release). The top panels show, from top left to bottom right, one-to-one relationships between stellar mass ($M_*$), SFR (averaged over the most recent $100\ \mathrm{Myr}$), dust attenuation ($A_V$), and mass-weighted age ($t_\mathrm{MWA}$). Red dotted lines show the one-to-one relations.}
    \label{fig:iyer_comparison}
\end{figure*}

\begin{figure}
    \centering
    \includegraphics[width=0.5\linewidth]{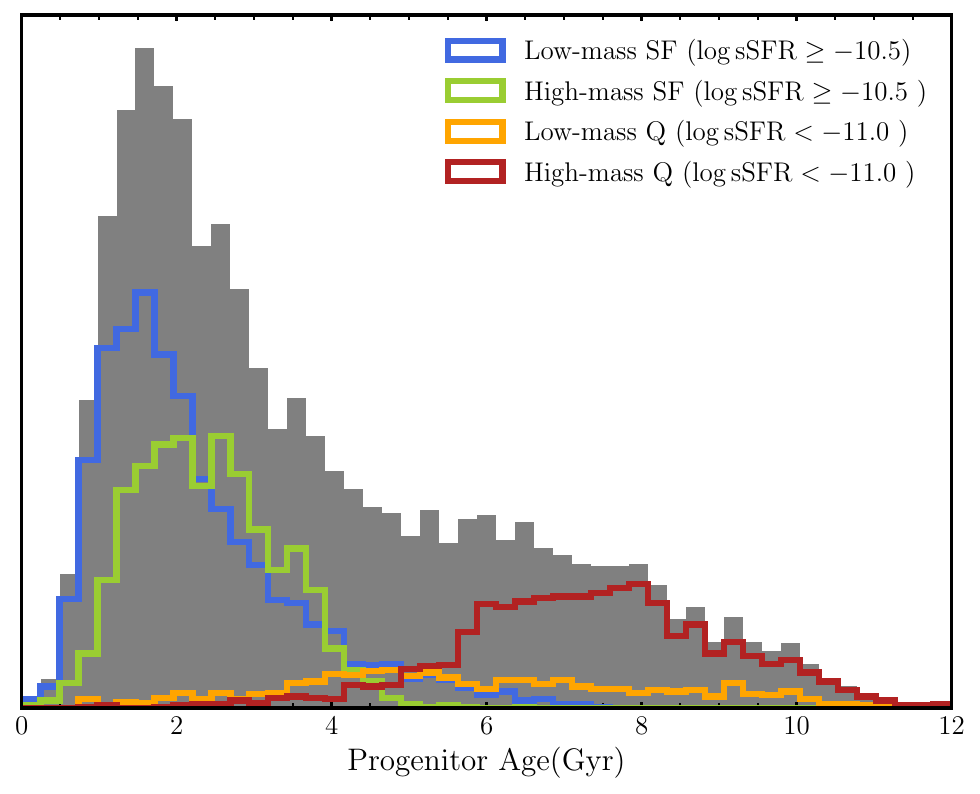}
    \caption{The stacked distribution of the expected progenitor age, $P(\mathbb{E}[t_\text{prog}])$, assuming a \cite{Leja2019} nonparametric SFH with a continuity prior. We find that the overall shape of the distribution is similar to that derived when assuming an \cite{Iyer_2019_nonparamSFH} SFH, and the mean remains largely unchanged at $3.8$~Gyr under the baseline DTD.}
    \label{fig:leja_age_histogram}
\end{figure}

\begin{figure}
    \centering
    \includegraphics[width=\linewidth]{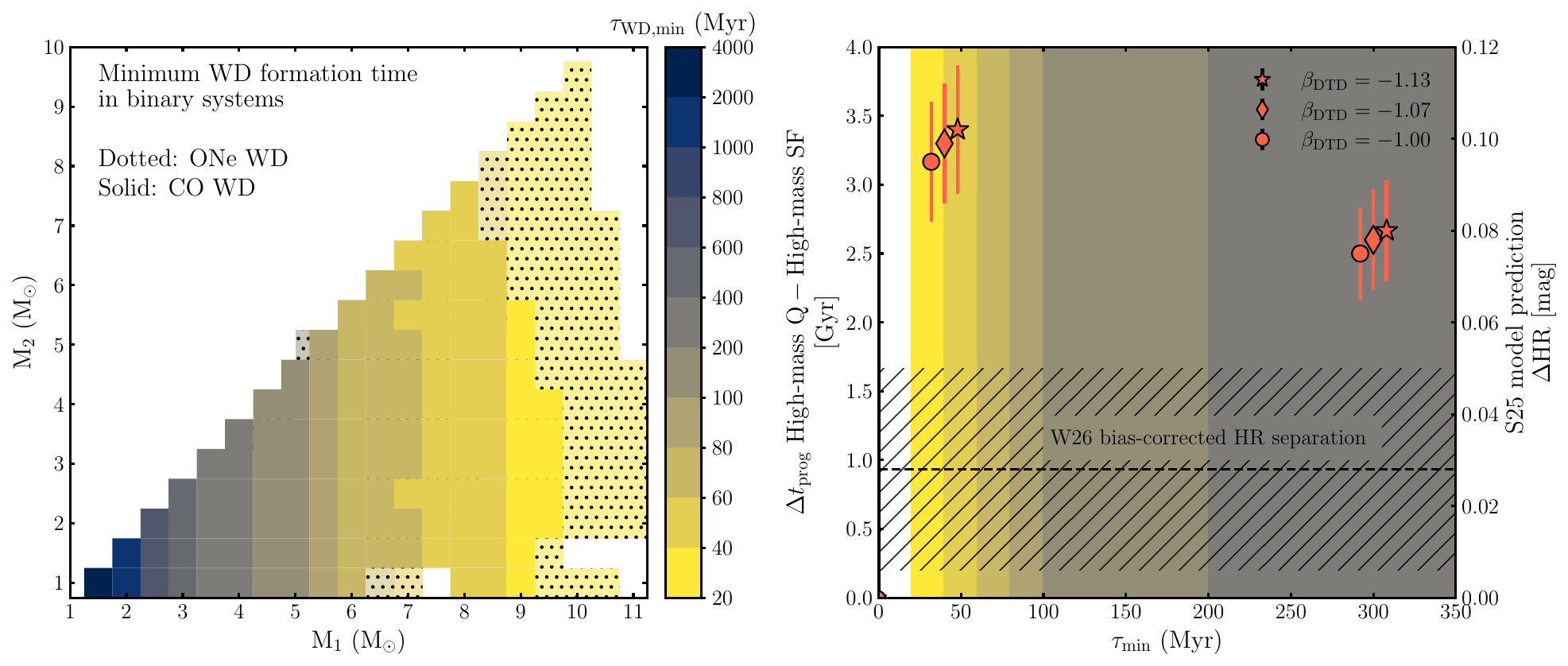}
    \caption{Sensitivity of our results discussed in Sec.~\ref{sec:age} to choice of DTD parameters. Binary evolution indicates a short cutoff time $t_\mathrm{min}\approx 40$ Myr. \textit{Left:} The minimum WD formation time (in Myr) from binary simulation. Main-sequence stars with large initial mass can become WDs within 30 Myr, and such systems can immediately yield an SN~Ia within a few tens of Myr with mass transfer from the companion. A larger minimum delay time of $t_{\text{min}}=300$~Myr requires that such early-formed WD systems \textit{never} produce SNe for another $\sim250$~Myr. \textit{Right:} The measured separation of mean progenitor ages between high-mass quiescent (Q) and high-mass star-forming (SF) hosts (see Sec.~\ref{sec:age_tests_S25}), as well as the corresponding predictions of the HR offset from the \Son model. The choice of $t_\mathrm{min}$ and the power-law slope $\beta_\mathrm{DTD}$ can affect the comparison to the observed HR separation (hatched region) by $\sim20$\%.}
    \label{fig:tau_min}
\end{figure}

\begin{figure*}
    \centering
    \includegraphics[width=1\linewidth]{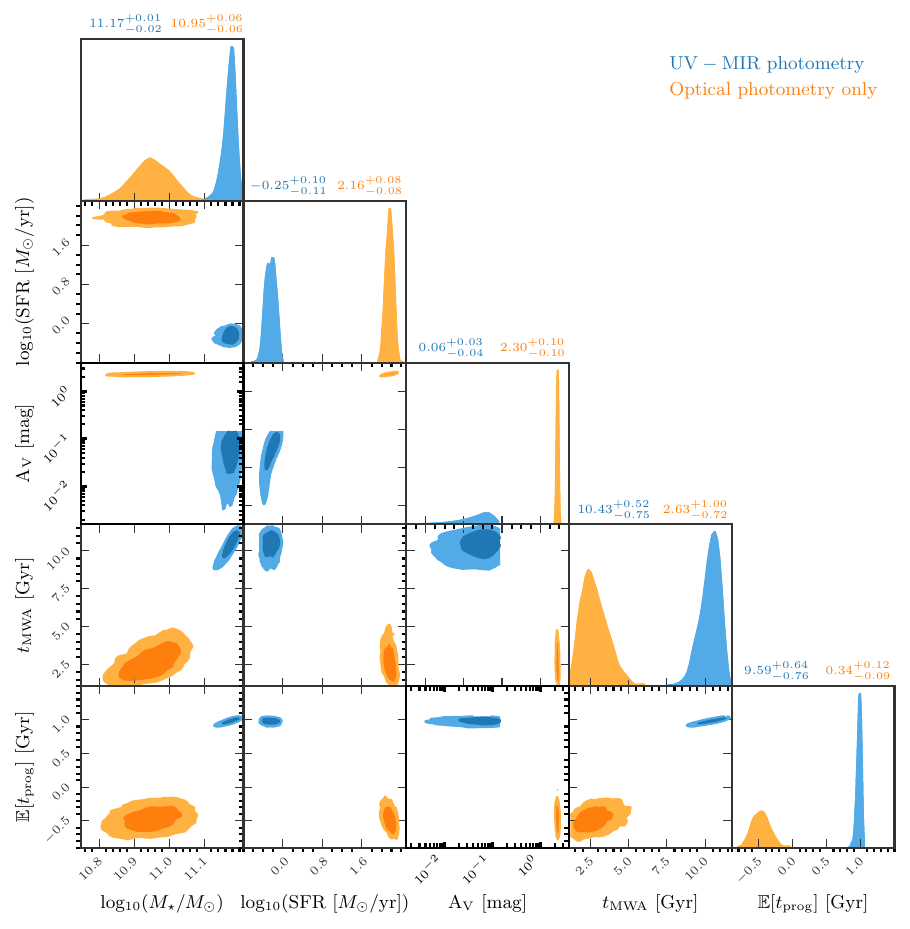}
    \caption{Corner plot showing the posterior probability distributions of the stellar population parameters derived from \texttt{Bagpipes} SED fitting of the host galaxy of SN~2024vjb. The blue contours correspond to fits using UV--MIR photometry, while the orange contours show results using only optical photometry. One-dimensional histograms (diagonal panels) and two-dimensional credible regions (off-diagonal panels) are displayed for (from left to right) stellar mass, SFR (averaged over the most recent $100\ \mathrm{Myr}$ of star-formation activity), dust attenuation, galaxy-mass-weighted age, and SN~Ia progenitor age. Median values and their uncertainties derived from the 16$^{\mathrm{th}}$ and 84$^{\mathrm{th}}$ percentiles are shown above each plot. Including UV--MIR data significantly tightens constraints on dust attenuation and SFR, and reduces degeneracies between $A_V$, SFR, and age relative to the optical-only fit.}
    \label{fig:2024vjb_corner}
\end{figure*}

In this work we adopt the Gaussian-process-based SFH formalism of \citet{Iyer_2019_nonparamSFH}, rather than the continuity-prior-based SFH of \citet{Leja2019} used in the TITAN DR1 data release (Tweddle et al., in prep.). As discussed in the main text, the primary motivation for this choice is to obtain a smoother, effectively continuous SFH representation, which avoids discretization effects when convolving with a DTD to infer SN progenitor-age distributions.

In fitting the \cite{Iyer_2019_nonparamSFH} within \texttt{Bagpipes}, we adopt identical assumptions and priors to those used in the TITAN DR1 analysis (Tweddle et al., in prep.; Table 3) for all other model components (i.e., a \citealt{Salim_2018_GSWLC2} dust model, the parameters of the \cite{Draine2007} polyaromatic hydrocarbon emission model, etc.). In our implementation, the SFH is described using five adaptive time bins, with star formation distributed between bins via a Dirichlet prior ($\alpha=1$), while allowing the total mass formed, metallicity, and overall SFR normalization to vary within broad bounds.

To assess the impact of this judicious choice of SFH on the derived parameters, Figure~\ref{fig:iyer_comparison} compares the resulting posterior median estimates of stellar mass, SFR, dust attenuation, and mass-weighted age obtained using the \cite{Iyer_2019_nonparamSFH} SFH to those derived using the \citet{Leja2019} continuity SFH. As illustrated in the figure, the inferred $M_*,\ \mathrm{SFR},\ \mathrm{and}\ A_V$ exhibit excellent agreement, with no systematic offsets and scatter consistent with the quoted uncertainties. We do, however, observe a modest shift in the inferred mass-weighted ages, with a best-fit relation of slope $1.18 \pm 0.01$ and intercept $-0.09 \pm 0.05$. This behavior is expected, as $t_\mathrm{MWA}$ is inherently sensitive to the detailed temporal structure of the SFH. Crucially, this shift does not alter the relative ordering of galaxies in age: systems identified as young (old) under the \citet{Leja2019} SFH generally remain systematically young (old) under the \cite{Iyer_2019_nonparamSFH} formalism.

Furthermore, in Figure~\ref{fig:leja_age_histogram} we present the distribution of $P(\mathbb{E}[t_\mathrm{prog}])$ inferred when adopting the nonparametric SFH reconstruction of \cite{Leja2019}. In this case, we find a mean progenitor age of $\langle t_\text{prog}\rangle=3.8$~Gyr, and the overall progenitor-age distribution is remarkably consistent with that obtained using our fiducial SFH modeling. Consequently, our inference of a comparatively young SN~Ia progenitor population does not appear to depend sensitively on the adopted SFH formalism, strengthening the case that the tension with strong progenitor-age evolution scenarios is physical rather than methodological.

We therefore conclude that our results are robust to the choice of nonparametric SFH, and that adopting the \cite{Iyer_2019_nonparamSFH} SFH formalism does not introduce any material differences in the inferred bulk galaxy or progenitor-age properties. The primary effect of this choice is instead to provide a smoother and better-behaved SFH for downstream analyses that depend sensitively on the temporal structure of star formation, such as the inference of SPADs.

\section{Justification of DTD Parameters} \label{sec:app_dtd}

\Wiseman show that the choice of DTD parameters could lead to differences in predicted progenitor ages. Owing to the strong sensitivity to recent star formation, the inferred progenitor ages are sensitive to the cutoff time $t_\mathrm{min}$, below which no SN is formed from the youngest stellar population. This is largely dictated by the formation time for a WD in a compact binary. 

We use SeBa, a fast binary evolution code, to explore the minimum WD formation time \citep{portegies_zwart_population_1996,nelemans_population_2001,toonen_supernova_2012}. The $\gamma\alpha$ prescription is followed for the common-envelope phase with $\alpha=4$, $\lambda=0.5$, and $\gamma=1.75$, as prescribed by \cite{toonen_supernova_2012}, which can influence systems with high initial masses and small separations that undergo a common-envelope phase. We simulate 23,100 binaries for 13.5 Gyr on a fixed grid with binary component masses of 1--12 $M_{\odot}$, binary separations of 50--$10^4$ $R_{\odot}$, zero eccentricity, and solar metallicity ($Z = 0.014$). The system is classified as having formed a WD if it is in a detached binary and the companion is yet to form a compact remnant. For a given initial mass of the two component stars, we calculate $\tau_{\mathrm{min}}$ by comparing WD formation times for different initial separations. 

The left panel of Figure~\ref{fig:tau_min} shows the result of a binary evolution code and demonstrates that the minimum formation time is $\sim30$--50 Myr for $\sim7$--9 $M_\odot$ stars. Such WDs with large companions can quickly produce SNe~Ia through mass transfer, supporting the choice of $t_\mathrm{min}\approx 40$ Myr from an astrophysical perspective. If one employs a larger minimum delay time for SNe~Ia, such as $t_\text{min} = 300$~Myr (following \citealt{Childress_2014_age} or \Son), early-forming systems, indicated by light shaded colors in Figure~\ref{fig:tau_min}, are required to wait for an additional $\sim250$~Myr (because \textit{no} SNe~Ia are allowed to occur on timescales shorter than $t_\text{min}$). 
This poses an astrophysical difficulty to a large $t_\text{min}$ DTD model, since 7--8 $M_\odot$ secondary stars can evolve out of the main sequence within $10^0$--$10^1$~Myr from the formation of the CO WD and initiate further mass transfer, setting up an environment for the CO-WD--helium-star channel for SNe~Ia.
This preference toward shorter $t_\text{min}$ is consistent with the literature in binary synthesis, particularly for CO-WD--He-star systems \citep[e.g.,][]{Wang_2009_COWD_HeStar,Claeys_2014_WD_HeStar,Yungelson_2000_SNRate_CosmicHistory}.

Varying this minimum delay time between 20~Myr and 300~Myr (right panel of Figure~\ref{fig:tau_min}) affects the predicted ages of the youngest SN~Ia progenitors; the resulting separation between young and old populations, arising respectively from star-forming and quiescent hosts, varies by $\sim20$\%. This level of variation does not affect our result, as the discrepancy between the \Son-model prediction based on this age difference and HR offsets reported by \Wiseman (hatched region) are systematically discrepant, especially considering the astrophysical difficulties with the $t_\text{min}=300$~Myr model shown by binary synthesis.

\section{\texttt{Bagpipes} Fits} \label{sec:app_bagpipes}

Figure~\ref{fig:2024vjb_corner} presents the full posterior distributions from the \texttt{Bagpipes} SED fitting of the host galaxy of SN~2024vjb. The blue contours show the results obtained when fitting the full UV--MIR photometric dataset, while the orange contours correspond to fits using only optical photometry. As discussed in the main text, the inclusion of MIR photometry significantly reduces degeneracies between $A_V$, SFR, and age, leading to a systematically older and more tightly constrained progenitor age compared to the optical-only fit.

\bibliography{main}{}
\bibliographystyle{aasjournal}

\end{document}